\documentclass[12pt]{article}
\usepackage{a41}

\usepackage{palatino}
\usepackage{xcolor}
\usepackage{amsmath}
\usepackage{booktabs}
\usepackage{microtype}
\usepackage{cite}
\usepackage{amsmath}
\usepackage{amssymb}
\usepackage{color}
\usepackage[rflt]{floatflt}
\usepackage{float}

\newcommand{\bbb}{\Bigg}
\newcommand{\bba}{\bigg}

\usepackage[rflt]{floatflt}
\usepackage{float}
\usepackage{slashed}
\def\b0{\beta_0}

\usepackage{array}

\usepackage[english]{babel}
\usepackage[latin1]{inputenc}
\usepackage[T1]{fontenc}
\usepackage{ae}

\usepackage{url}

\usepackage{amsmath, amsthm, amssymb}
\newtheorem{thm}{Theorem}[section]

\newtheorem{definition}[thm]{Definition}

\newcommand{\ep}{\varepsilon}

\usepackage{rotating}

\usepackage{graphicx}
\newcounter{mmacnt}
\def\restartmma{\setcounter{mmacnt}{0}}
\restartmma \catcode`|=\active
\def|#1|{\mathrm{#1}}
\catcode`|=12
\newenvironment{mma}{
 \par\smallskip
 \catcode`|=\active
 \parskip=0pt\parindent=0pt % locally
 \small
 \def\In##1\\{%
\def\linebreak{\hfill\break\null\qquad}%
\refstepcounter{mmacnt}
\hangindent=2.5em\hangafter=0
\leavevmode
\llap{\tiny\sffamily n[\arabic{mmacnt}]:=\kern.5em}%
\mathversion{bold}\footnotesize$\displaystyle##1$\normalsize
\mathversion{normal}\par
 }%
 \def\Print##1\\{%
\def\linebreak{\hfill\break}%
\hangindent=2.5em\hangafter=0
\leavevmode ##1\par}%
 \def\Out##1\\{%
\def\linebreak{$\hfill\break\null\hfill$}%
\kern\abovedisplayskip\par
\hangindent=2.5em\hangafter=0
\leavevmode
\llap{\tiny\sffamily Out[\arabic{mmacnt}]=\kern.5em}
\footnotesize$\displaystyle##1$\normalsize\hfill\null\par
\kern\belowdisplayskip
 }%
 \def\Warning##1##2\\{%
\def\linebreak{\hfill\break}%
\hangindent=2.5em\hangafter=0
\leavevmode
{\scriptsize##1 : ##2}\par}%
}{%
 \par\smallskip
}

\usepackage{color}

\newenvironment{fshaded}{%
\MakeFramed {\FrameRestore}
}%
{\endMakeFramed}

\def\b0{\beta_0}

\def\Gp0{{\Gamma^{'}_0}}
\def\Gp1{{\Gamma^{'}_1}}
\def\Gp2{{\Gamma^{'}_2}}

\allowdisplaybreaks[4]

\begin{document}
\setlength{\baselineskip}{0.515cm}

\sloppy
\thispagestyle{empty}
\begin{flushleft}
DESY 20--025
\hfill 
%%{\tt arXiv:2002.xxxxx[gr-qc]}
\\
DO--TH 20/01\\
SAGEX--20--03\\
\end{flushleft}

\mbox{}
\vspace*{\fill}
\begin{center}

{\Large\bf Fourth post-Newtonian Hamiltonian  dynamics of}

\vspace*{3mm}
{\Large\bf two-body systems from an effective field theory approach}

\vspace{3cm} \large
{\large J.~Bl\"umlein$^a$, A.~Maier$^a$, P.~Marquard$^a$, and G.~Sch\"afer$^b$}

\vspace{1.cm}
\normalsize
{\it $^a$Deutsches Elektronen--Synchrotron, DESY,}\\
{\it   Platanenallee 6, D--15738 Zeuthen, Germany}

\vspace*{2mm}
{\it $^b$Theoretisch-Physikalisches Institut, Friedrich-Schiller-Universit\"at, \\
Max-Wien-Platz 1, D--07743 Jena, Germany}\\

%%\today

\end{center}
\normalsize
\vspace{\fill}
\begin{abstract}
\noindent
We calculate the motion of binary mass systems in gravity up to the fourth post--Newtonian order. 
We use momentum expansions within an effective field theory approach based on Feynman amplitudes 
in harmonic coordinates by applying dimensional regularization. We construct the canonical transformations 
to ADM coordinates and to effective one body theory (EOB) to compare with other approaches. We show that 
intermediate poles in the dimensional regularization parameter $\ep$  vanish in the observables and 
the classical theory is not renormalized. The results are illustrated for a series of observables for 
which we agree with the literature.
\end{abstract}

\vspace*{\fill}
\noindent
% \numberwithin{equation}{section}
%%%%%%%%%%%%%%%%%%%%%%%%%%%%%%%%%%%%%%%%%%%%%%%%%%%%%%%%%%%%%%%%%%%%%%%%%%%%%%%%%%%%%%%%%%%%%%%%%%%%%%%%%%%%%%%%%%%%%%%%%%%%%%%%%%%
\newpage

%--------------------------------------------------------------------------------------------------------
\section{Introduction}
\label{sec:1}
%--------------------------------------------------------------------------------------------------------

\vspace*{1mm}
\noindent
Precise predictions for observables describing the merging of two heavy astrophysical objects like black holes or 
neutron stars are very important \cite{TheLIGOScientific:2016pea}. In particular the spiral-in phase of these 
processes can be described by analytic calculations. Methods of non-relativistic effective field theory 
\cite{EFT,Kol:2007bc,Gilmore:2008gq,Foffa:2011ub,Foffa:2019rdf,Foffa:2019yfl,Foffa:2019hrb,Blumlein:2019zku} allow
to calculate the equations of motion of a binary mass system within the post--Newtonian (PN) approach. We will 
limit ourselves to the non-spinning case in the following. Until now the corrections have been computed to the 4th 
post--Newtonian order \cite{Kol:2007bc,Gilmore:2008gq,Foffa:2011ub,Foffa:2019rdf,Foffa:2019yfl}. There are first 
corrections due to the static potential at 5PN \cite{Foffa:2019hrb,Blumlein:2019zku}, see also \cite{Foffa:2019eeb}. 
The corrections up to 4PN were derived by using other methods, cf.~ Refs.~\cite{Damour:2014jta,Bernard:2017ktp},
before. Very recently first partial results up to the 5th post--Newtonian order have been obtained in 
\cite{Bini:2019nra}. In the post--Newtonian approach one retains all terms in the velocity being of the same order 
as the pure potential terms by the virial theorem \cite{CLAUSIUS}. Recently also important progress has been made 
in the post-Minkowskian (PM) approach reaching 3PM cf.~\cite{Bern:2019crd,Damour:2019lcq}, see also~\cite{PM}.

In this paper we present the corrections up to 4PN obtained using effective field theory methods. Here we 
apply dimensional regularization in calculating the Feynman integrals. At intermediary steps the dimensional
parameter $D  = 4 - 2\ep$ occurs. We perform multiple comparisons to results in the literature and find 
agreement. Starting at 3PN singular contributions of $O(1/\ep)$ occur when working in harmonic coordinates.
From 4PN onward  singularities of $O(1/\ep)$ also result from the tail terms, cf. also \cite{Foffa:2011ub,Foffa:2019rdf}.
By applying canonical transformations one may map the Hamiltonian in harmonic coordinates into a class of pole--free 
Hamiltonians at 3PN, to which
the ADM and EOB Hamiltonians belong. This is also the case 
at 4PN when accounting for the tail terms before. Therefore, the classical observables are free of the 
intermediary regularization parameter. 

The paper is organized as follows. In Section~\ref{sec:2} we describe the calculation method. We outline the general
framework and present the results up to 4PN in harmonic coordinates by deriving the effective Hamiltonian using the 
Hamiltonian formalism \cite{Schafer:2018kuf} in the center of momentum frame. The cancellation of the pole 
contributions is discussed. We also present an associated pole--free Hamiltonian obtained after applying a canonical 
transformation. At 3PN one also obtains a $\ln(r)$ free Hamiltonian, with $r = |\vec{x}_1 - \vec{x}_2|$.  In 
Section~\ref{sec:3} we construct the canonical
transformation to ADM coordinates, to compare with these results. In Section~\ref{sec:4} we study the canonical 
transformation to EOB coordinates, which are widely used in the literature and present the EOB Hamiltonian up to 4PN 
in explicit form. We present numerical results for the energy of the innermost stable orbit and the angular momentum for 
circular coordinates in Section~\ref{sec:5}. Section~\ref{sec:6} contains the conclusions. In an appendix we present 
the general $D$-dimensional Hamiltonian in harmonic coordinates up to $O(\ep)$ in explicit form, which is 
also important for higher post--Newtonian calculations.
%--------------------------------------------------------------------------------------------------------
\section{Calculation Method}
\label{sec:2}
%--------------------------------------------------------------------------------------------------------
%--------------------------------------------------------------------------------------------------------
\subsection{The general framework}
\label{sec:21}
%--------------------------------------------------------------------------------------------------------

\vspace*{1mm}
\noindent
The main steps of our calculation have already been described in Ref.~\cite{Blumlein:2019zku} calculating the static
potential to 5PN ab initio. Now we add the different velocity contributions up to 4PN. Starting from the Einstein--Hilbert
Lagrangian, one parameterizes the metric $g_{\mu\nu}$ as proposed in \cite{Kol:2007bc} in terms of one scalar, 
a three-component vector and 
a six-component tensor field. Within this parameterization one can derive the contributing Feynman rules using the path 
integral.\footnote{The Feynman rules used in the present calculation are to lengthy to be presented here. They will be 
given elsewhere.} The corresponding Feynman diagrams are generated using {\tt QGRAF} \cite{Nogueira:1991ex}, after providing 
the corresponding Feynman rules.  We consider only the classical contributions and calculate the effective two--body potential.
Here the necessary velocity contributions range up to $O((G_N \overline{M}/r)^k (v^2)^{n-k+1})$, with $G_N$ Newton's 
constant, $\overline{M}$ a mass scale, $r$ the 
distance of the point masses and $v \in \{v_1, v_2\}$ their velocities. We will set $c = 1$ in many places and keep it 
only when needed as an order parameter. The Lorentz algebra is carried out using {\tt Form} 
\cite{FORM} and the 
reduction to master integrals is performed using the code {\tt Crusher} \cite{CRUSHER}. The contributing master integrals 
are known from the calculation of the static potential already, cf. \cite{Blumlein:2019zku} and references therein.
One first obtains a Lagrange function of $m$th order containing also the accelerations $a_i$ and time derivatives thereof.

In Table~\ref{TAB1} we give an overview on the complexity of the calculation up to 4PN and give the numbers of generated
diagrams, the number of non factorizing diagrams, those with no world-line loops and no tadpoles by loop order. The initial 
number of 25324 diagrams reduces to 9266 contributing to the present result. Here we do not sort them in addition into 
equivalence 
classes and 
do thus determine the final combinatorial factors by the calculation itself.
%--------------------------------------------------------------------------------------------------------
\begin{table}[H]\centering
\renewcommand{\arraystretch}{1.3}
\setlength{\tabcolsep}{11pt}
\begin{tabular}{rrrrrr@{\quad}r}
\toprule
\multicolumn{1}{c}{ }            &
\multicolumn{1}{c}{{\tt QGRAF}}         &
\multicolumn{1}{c}{non fact.}     &
\multicolumn{1}{c}{no WL loops}     &
\multicolumn{1}{c}{no tadpoles}     &
\multicolumn{1}{c}{\# Diag. \cite{Foffa:2019rdf,Foffa:2012rn}}  &
\multicolumn{1}{c}{\# MI} \\
\midrule
  0  &      3  &      3 &      3 &     3 & 3   & 0\\
  1  &     70  &     70 &     70 &    70 & 23  & 1\\
  2  &   1770  &   1770 &   1770 &  1468 & 212 & 1\\
  3  &  13400  &   9792 &   9482 &  5910 & 317 & 1\\
  4  &  10081  &   5407 &   4685 &  1815 & 50  & 4(3)\\
\bottomrule
\end{tabular}
\caption[]{\small \sf Numbers of contributing diagrams at the different loop levels
and master integrals (MI). The numbers in brackets denote the number of master integrals which occur
during the reduction but do not contribute to the potential. In the  next-to-last column the number of
diagrams of the respective equivalence classes are given according to \cite{Foffa:2012rn,Foffa:2019rdf}.}
\label{TAB1}
\renewcommand{\arraystretch}{1.0}
\end{table}
%--------------------------------------------------------------------------------------------------------
\noindent
The computation time for the complete project from Feynman diagram generation, IBP reduction to the final results
amounts to a few hours using an {\tt Intel(R) Core(TM) i7-8650U CPU}. 

The present results and those from \cite{Gilmore:2008gq,Foffa:2012rn,Foffa:2019rdf}
are given in form of an $m$th
order Lagrange-density, from which one may derive the associated equation of motion \cite{SCHMUTZER}
%--------------------------------------------------------------------------------------------------------
\begin{eqnarray}
\label{eq:EOM}
\delta S = \int d^D x \left\{
\sum_{k=0}^m (-1)^{k+1} \frac{d^k}{dt^k} \left[\frac{\partial {\cal L}}
  {\partial U_{\Omega|{\tiny \underbrace{t..t}_{k}}}}
\right]\right\} \delta U_\Omega = 0,
\end{eqnarray}
%--------------------------------------------------------------------------------------------------------
with ${\cal L}$ the Lagrangian density and $U_\Omega \in \{x_1,x_2\}$, the coordinates of the two point masses and
$U_{\Omega|t} = \partial_t U_\Omega$.
The present results, although widely different in form and probably based on different Feynman rules, can be compared
using the equation of motion (\ref{eq:EOM}) finding agreement with the corresponding equation of motion for the
Lagrangians given in \cite{Gilmore:2008gq,Foffa:2012rn,Foffa:2019rdf}. This applies to the potential contributions. In 
addition the tail terms have to be considered, cf.~\cite{Foffa:2019rdf,Porto:2017dgs,Galley:2015kus}. 

The next step is to eliminate 
the accelerations and their time derivatives. For this we use double zero insertions \cite{Damour:1985mt}, up to 
linear terms in the acceleration. These are eliminated by a coordinate shift \cite{DW,Damour:1985mt,Damour:1990jh} 
implying 
the contribution of  variational derivatives of the Lagrangian and a total time 
derivative. This procedure is applied to the $D$-dimensional Lagrangian for all post--Newtonian orders.
The first order Lagrangian density is then obtained in somewhat different coordinates from harmonic 
coordinates \cite{Damour:1990jh}. Finally, we perform the Legendre transformation to obtain the Hamiltonian.
%--------------------------------------------------------------------------------------------------------
\subsection{Remarks on the 4PN Tail Term}
\label{sec:22}
%--------------------------------------------------------------------------------------------------------

\vspace*{1mm}
\noindent
The quadrupole structure of the tail term at 4PN results from the leading term in the multipole expansion of the retarded 
radiation field
from linearized Einstein gravity, cf.~\cite{Marchand:2017pir},
%---------------------------------------------------------------------------------------------------------
\begin{eqnarray}
h^{\mu\nu} = |g|^{1/2}g^{\mu\nu} - \eta^{\mu\nu},
\end{eqnarray}
%---------------------------------------------------------------------------------------------------------
with $\eta^{\mu\nu}$ the Minkowski metric and $|g|$ the modulus of the determinant of the metric tensor. Let $L$ be the 
multi--index $L = 
i_1,i_2,...,i_l$, then
the components of $h^{\mu\nu}$ are given by
%---------------------------------------------------------------------------------------------------------
\begin{eqnarray}
h^{00} &=& - \frac{4G_N}{c^2}\sum_{l=0}^{\infty}\frac{(-1)^l}{l!}\partial_L\tilde{I}_L,
\\
h^{0i} &=& \frac{4G_N}{c^3}\sum_{l=1}^{\infty}\frac{(-1)^l}{l!}\partial_{L-1}\dot{\tilde{I}}_{iL-1},
\\
h^{ij} &=& -\frac{4G_N}{c^4}\sum_{l=2}^{\infty}\frac{(-1)^l}{l!}\partial_{L-2}\ddot{\tilde{I}}_{ijL-2},
\end{eqnarray}
%---------------------------------------------------------------------------------------------------------
where
%---------------------------------------------------------------------------------------------------------
\begin{eqnarray}
\tilde{I}_L(t,r) = \frac{\Gamma((D-3)/2)}{\pi^{(D-3)/2}~ r^{D-3}}\int_1^{+\infty}dz ~\gamma_{(2-D)/2}(z) ~I_L(t-zr/c)
\end{eqnarray}
%---------------------------------------------------------------------------------------------------------
and
%---------------------------------------------------------------------------------------------------------
\begin{eqnarray}
\gamma_{(2-D)/2}(z) = \frac{2\pi^{1/2}}{ \Gamma((4-D)/2)\Gamma((D-3)/2)} (z^2 - 1)^{(2-D)/2}.
\end{eqnarray}
%---------------------------------------------------------------------------------------------------------
The dots are Newton's notation for the time-derivative. $I_L(t)$ are the mass-type multipole moments in Cartesian 
coordinates in symmetric-trace-free (STF) tensor representation. We do not consider the current-type multipoles, since 
they do not contribute at the level of 4PN.
 
The leading order tail contribution results from the scattering of the $I_{ij}(t)$-generated quadrupole radiation field off 
the static field $h^{00} = - \frac{4G}{c^2}\tilde{I}_0$ from the monopole source $l=0$,  
%---------------------------------------------------------------------------------------------------------
\begin{eqnarray}
\tilde{I}_0 = M\frac{\Gamma((D-3)/2)}{\pi^{(D-3)/2}}r^{3-D},
\end{eqnarray}
%---------------------------------------------------------------------------------------------------------
where $M = m_1 + m_2$ denotes the total mass of the radiating system. The source representation of $I_{ij}(t)$ 
in the case of a binary system reads in its center-of-mass system  
%---------------------------------------------------------------------------------------------------------
\begin{eqnarray}
I_{ij} = \mu \left(r_i r_j - \frac{\delta_{ij}r^2}{D-1}\right),
\end{eqnarray}
%---------------------------------------------------------------------------------------------------------
with $\mu = m_1 m_2/M$. More details are given in \cite{Marchand:2017pir,Bernard:2017bvn}.
A derivation of the tail-term in ADM-coordinates was given in  \cite{GS:1990,Damour:2016abl}.

Since the derivation of the 4PN tail term is not thoroughly unique in the literature due to the use of different 
regularization schemes or combinations thereof, cf.~\cite{Blanchet:1987wq,Foffa:2011np,Damour:2014jta,Jaranowski:2015lha,
Galley:2015kus,Bernard:2017bvn,Foffa:2019rdf}, we add a few clarifying remarks in the present treatment based on Feynman 
amplitudes. We follow Ref.~\cite{Foffa:2019rdf}, Eq.~(23), performing the calculation in $D$ dimensions. 
The gravitational constant is taken in $D$ dimensions, $G_N \rightarrow G_N \mu_1^{-2\ep}$, with a
mass scale $\mu_1$ which reappears in
%--------------------------------------------------------------------------------------------------------
\begin{eqnarray}
\label{eq:r0}
r_0 = \frac{e^{-\gamma_E/2}}{2 \sqrt{\pi} \mu_1}
\end{eqnarray}
%--------------------------------------------------------------------------------------------------------
and $\gamma_E$ denotes the Euler--Mascheroni constant.\footnote{
In other calculations \cite{Jaranowski:2015lha} the regularization has been performed using the Hadamard 
symbol \cite{LUETZEN,HADAMARD}.}

Adjusting the expression from \cite{Foffa:2019rdf} for the different choice of regulator $-2 \varepsilon = 
\varepsilon_{\mathrm{FS}}$   the representation for the tail contribution to the action $S$ is given by 
%--------------------------------------------------------------------------------------------------------
\begin{equation}
  \label{eq:PM6a}
  \delta S _{\mathrm{tail}} = -\frac{G_N^2 M }{5}\int_{-\infty}^\infty \frac{dk_0}{2 \pi} \,k_0^6 
\left ( -\frac{1}{2 \varepsilon} - \frac{41}{30} + \ln \left( \frac{ k_0^2 e^{\gamma_E}}{\pi \mu_1^2} \right) \right ) 
Q_{ij}(k_0)Q^{ij}(-k_0),
\end{equation}
%--------------------------------------------------------------------------------------------------------
where $Q(k_0)$ denotes the Fourier transform of the  quadrupole moment $I_{ij}$
%--------------------------------------------------------------------------------------------------------
\begin{equation}
  \label{eq:PM18}
Q_{ij}(k_0) = {\cal F}[I_{ij}](k_0) \equiv \int_{-\infty}^\infty  dt\, e ^{i k_0 t} I_{ij}(t)   \,.
\end{equation}
%--------------------------------------------------------------------------------------------------------
To convert this expression to the time domain we have to make use of
the well--known  property of the Fourier transform \cite{VLADIMIROV}
%--------------------------------------------------------------------------------------------------------
\begin{equation}
  \label{eq:PM19}
  k_0 {\cal F}[I_{ij}](k_0) = -i  {\cal F}[\dot I_{ij}](k_0) \,.
\end{equation}

We finally  obtain the following  expression\footnote{Here and in the following summation over equal indexes is 
understood.}
%--------------------------------------------------------------------------------------------------------
\begin{eqnarray}
  \label{eq:PM6}
  \delta S _{\mathrm{tail}} &=& \frac{G_N^2 M }{5}\int_{-\infty}^{+\infty} dt  \Bigg \{  \left ( \frac{1}{2 \varepsilon} + 
\frac{41}{30} \right ) \left(\dddot {I_{ij}}(t) \right)^2   \nonumber\\
&&  +  \dddot {I_{ij}}(t) \int_0^{+\infty} d\tau  \ln \left( \frac{ \tau }{\tau_0} \right ) \bigg ( \ddddot {I_{ij}} (t- 
\tau) - \ddddot {I_{ij}} (t+ \tau) \bigg )\Bigg \}
\end{eqnarray}
%--------------------------------------------------------------------------------------------------------
and $\tau_0$ is given by
%--------------------------------------------------------------------------------------------------------
\begin{eqnarray}
\tau_0 =   2 r_0.
\end{eqnarray}
%--------------------------------------------------------------------------------------------------------
%The quadrupole moment $I_{ij}(t)$ is given by 
%--------------------------------------------------------------------------------------------------------
%\begin{equation}
%  \label{eq:PM16}
%  I_{ij}(t) = \sum_{k=1,2} m_k \left ( x_k^i x_k^j  - \frac{1}{d} |\vec x_k|^2 \right )
%\end{equation}
%--------------------------------------------------------------------------------------------------------
The square of the third time derivative of $I_{ij}$ evaluates to 
%--------------------------------------------------------------------------------------------------------
\begin{align}
  \label{eq:PM17}
  &(\dddot I_{ij}(t))^2 = \frac{G_N^2 M^2}{r^4} \Bigg [-\frac{88}{3} (n.p)^2+32 p^2 \nonumber \\
&+\varepsilon  \left(\left(128 p^2-\frac{352}{3} (n.p)^2\right) \ln \left(\frac{r}{r_0}\right)+\frac{1672}{9} (n.p)^2-192 
p^2\right)\Bigg] \,.
\end{align}
%--------------------------------------------------------------------------------------------------------
Note, that we need to keep terms  ${\cal O}(\varepsilon)$ that are
multiplied with the pole term in $\delta S_{\mathrm{tail}}$.   

On circular orbits, setting $p.n  = 0$, the $\tau$ integration in Eq.~(\ref{eq:PM6}) can be
performed analytically and one obtains 
%--------------------------------------------------------------------------------------------------------
\begin{equation}
  \label{eq:PM20}
  \delta S _{\mathrm{tail}} = \frac{G_N^2 M }{5} \int_{-\infty}^{+\infty} dt \left(\dddot
    I_{ij}(t) \right)^2 \Bigg \{    \frac{1}{2 \varepsilon} + 
    \frac{41}{30}   - 2 \big ( \gamma_E +  \ln{(2 \Omega \tau_0})
    \big)  \Bigg \} \,,
\end{equation}
%--------------------------------------------------------------------------------------------------------
with the angular frequency $\Omega$ given by
%--------------------------------------------------------------------------------------------------------
\begin{eqnarray}
\label{eq:OMEG}
\Omega = \frac{dE_{\rm circ}}{dJ},
\end{eqnarray}
%--------------------------------------------------------------------------------------------------------
with $E_{\rm circ}$ the energy and $J$ the angular momentum at the orbit. 
For later use we introduce the variable $x$
%--------------------------------------------------------------------------------------------------------
\begin{eqnarray}
\label{eq:x}
x = (G_N M \Omega)^{2/3}.
\end{eqnarray}
%--------------------------------------------------------------------------------------------------------

In observables on circular orbits only the combination
%--------------------------------------------------------------------------------------------------------
\begin{equation}
  \label{eq:PM21}
  \ln \left ( \frac{r}{r_0} \right) + \ln(2 \Omega \tau_0) = \ln(4
\Omega r) = \frac{1}{2} \ln( 16 x)
\end{equation}
%--------------------------------------------------------------------------------------------------------
appears, which unambiguously gives rise to the well--known result in
the literature without the need to manually choose a particular
scale. 

Aside from this logarithm, logarithms of the type $\ln(r/r_0)$ remain in the Hamiltonian in harmonic 
coordinates. They will be treated in Section~\ref{sec:24}.
%--------------------------------------------------------------------------------------------------------
\subsection{The results in harmonic coordinates}
\label{sec:23}
%--------------------------------------------------------------------------------------------------------

\vspace*{1mm}
\noindent
In the following we list the different contributions to the Hamiltonian resulting from the first order Lagrange-function
derived in Section~\ref{sec:21} up to 4PN, which we write for center of mass system coordinates, normalizing the 
momentum
$p$ by $\mu$. The coordinates are close to harmonic coordinates, modified due to shifts to eliminate the 
accelerations. 
We normalize the Hamiltonian $H$ by $\hat{H} = H/M -1$, and refer to the parameter $\nu = \mu/M$.
Furthermore, we use $u = G_N M/r$. We account for the $O(\ep)$ terms up to 1PN since they play a role 
in the transformation to other coordinates discussed later. The full $O(\ep)$ up to 4PN is displayed in  
appendix~\ref{sec:A}.

One obtains 
%--------------------------------------------------------------------------------------------------------
\begin{eqnarray}
\label{eq:H0}
  \hat H_{\mathrm{N}} &=& \nu \Bigg \{ 
\frac{p^2}{2}
-u +
\ep
\bigg(
        1-2 \ln \left(
                \frac{r}{r_0}\right)\bigg) u 
\Bigg \}, \\
%-------
\hat H_{\mathrm{1PN}} &=&\nu \Bigg \{ 
\frac{1}{8} (-1+3 \nu ) p^4
+\frac{u^2}{2}
+ \bigg(
        \bigg(
                -\frac{3}{2}-\frac{\nu }{2}\bigg) p^2
        -\frac{1}{2} \nu  (p.n)^2
\bigg) u
\nonumber\\ &&
+\varepsilon  \bigg[
        \bigg(
                p^2 \bigg(
                        -\frac{1}{2}
                        +\frac{\nu }{2}
                        +(-3-\nu ) \ln \left(
                                \frac{r}{r_0}\right)
                \bigg) u
                +\bigg(
                        \frac{3 \nu }{2}
                        -\nu  \ln \left(
                                \frac{r}{r_0}\right)
                \bigg) (p.n)^2
        \bigg)
\nonumber\\ &&
        + \bigg(
                -1+2 \ln \left(
                        \frac{r}{r_0}\right)\bigg) u^2
\bigg]
\Bigg \}, 
\\
%-------
\hat H_{\mathrm{2PN}} &=& \nu \bbb \{ \frac{1}{16} \big(
        1-5 \nu +5 \nu ^2\big) p^6
+\frac{1}{4} (-2-\nu ) u^3
+ \bba(
        \frac{1}{4} (11+15 \nu ) p^2
        +\frac{1}{2} (-1+6 \nu ) (p.n)^2
\bba) u^2
\nonumber\\&&
+\bba(
        \frac{1}{8} \big(
                5-22 \nu -3 \nu ^2\big) p^4
        -\frac{1}{4} (-1+\nu ) \nu  p^2 (p.n)^2
        -\frac{3}{8} \nu ^2 (p.n)^4
\bba) u
\bbb \}, \\
%-----
\hat H_{\mathrm{3PN}} &=& \nu \bbb \{
\frac{5}{128} \big(
        -1+7 \nu -14 \nu ^2+7 \nu ^3\big) p^8
+\frac{1}{576} \big(
        216-10232 \nu +135 \nu  \pi ^2\big) u^4
\nonumber\\&&
+ \bba(
        \frac{1}{72} p^2 \big(
                -306+643 \nu 
-108 \nu ^2-63 \nu  \pi ^2\big)
        +\frac{1}{24} \big(
                36-2026 \nu +294 \nu ^2
\nonumber\\ &&
+63 \nu  \pi ^2\big) (p.n)^2
\bba) u^3
+ \bba(
        \frac{1}{16} (-1+8 \nu ) (29+12 \nu ) p^4
        +\frac{1}{4} \big(
                1-36 \nu -36 \nu ^2\big) p^2 (p.n)^2
\nonumber\\ &&        
+\frac{1}{3} \nu  (7+69 \nu ) (p.n)^4
\bba) u^2
+ \bba(
        -\frac{1}{16} \big(
                7-45 \nu +62 \nu ^2+5 \nu ^3\big) p^6
        -\frac{1}{16} \nu  \big(
                3-11 \nu +3 \nu ^2\big) 
        \nonumber\\&&
\times p^4 (p.n)^2
-\frac{3}{16} (-1+\nu ) \nu ^2 p^2 (p.n)^4
        -\frac{5}{16} \nu ^3 (p.n)^6
\bba) u
+\frac{1}{\varepsilon} \Biggl[ \frac{17 \nu  u^4}{6}
+ \Biggl(
        -\frac{17 \nu  p^2}{6}
\nonumber\\&&
+\frac{17}{2} \nu  (p.n)^2
\Biggr) u^3  \Biggr]
+  \Big[ \frac{68 \nu  u^4}{3}
+ \big(
        -17 \nu  p^2
        +51 \nu  (p.n)^2
\big) u^3  \Big] \ln \left(
        \frac{r}{r_0}\right)
\bbb\}, 
\\
%-----
\label{eq:H4}
\hat{H}_{4\rm PN} &=& \hat{H}_{4\rm PN}^{\rm loc} +  \hat{H}_{4\rm PN}^{\rm tail},
\\
%--------
\hat H_{\mathrm{4PN}}^{\rm loc} &=& \nu \bbb \{
\bba(
        \frac{7}{256}-\frac{63 \nu }{256}+\frac{189 \nu ^2}{256}-\frac{105 \nu ^3}{128}+\frac{63 \nu ^4}{256}\bba) p^{10}
\nonumber\\
&&+\bba(
        -\frac{3}{8}-\frac{123833 \nu }{1200}+\frac{33347 \nu ^2}{400}+\frac{79 \nu  \pi ^2}{64}-\frac{59 \nu ^2 \pi ^2}{8}\bba) u^5
\nonumber\\
&&
+ \bba[
        p^2 \bba(
                \frac{95}{16}+\frac{115733 \nu }{2880}-\frac{1223723 \nu ^2}{7200}+\frac{643 \nu  \pi ^2}{128}+\frac{1419 \nu ^2 \pi ^2}{128}\bba)
\nonumber\\
&&        +\bba(
                -\frac{11}{4}+\frac{73801 \nu }{1600}+\frac{953891 \nu ^2}{7200}-\frac{4429 \nu  \pi ^2}{192}-\frac{4033 \nu ^2 \pi ^2}{128}\bba) (p.n)^2
\bba] u^4
\nonumber\\
&&+ \bba[
        p^4 \bba(
                \frac{65}{16}-\frac{94439 \nu }{800}+\frac{319789 \nu ^2}{14400}+\frac{205 \nu ^3}{32}+\frac{1091 \nu  \pi ^2}{1024}-\frac{217 \nu ^2 \pi ^2}{64}\bba)\nonumber\\
&&
        +(p.n)^4 \bba(
                -\frac{6695 \nu }{32}-\frac{200369 \nu ^2}{320}-\frac{333 \nu ^3}{32}+\frac{4395 \nu  \pi ^2}{1024}+\frac{345 \nu ^2 \pi ^2}{128}\bba)
\nonumber\\
&&        +p^2 (p.n)^2 \bba(
                -\frac{5}{4}+\frac{294477 \nu }{800}+\frac{167173 \nu ^2}{1200}+\frac{11 \nu ^3}{16}-\frac{2955 \nu  \pi ^2}{512}+\frac{1095 \nu ^2 \pi ^2}{128}\bba)
\bba ] u^3
\nonumber\\
&&+  \bba[
        \bba(
                \frac{55}{32}-\frac{667 \nu }{64}+\frac{1217 \nu ^2}{64}-\frac{89 \nu ^3}{64}\bba) p^6
        +\bba(
                -\frac{3}{16}-\frac{99 \nu }{16}+\frac{733 \nu ^2}{16}
\nonumber\\
&& 
+\frac{3189 \nu ^3}{64}\bba) p^4 (p.n)^2
       +\bba(
                -\frac{79 \nu }{192}-\frac{4737 \nu ^2}{64}-\frac{7511 \nu ^3}{96}\bba) p^2 (p.n)^4
        +\bba(
                \frac{487 \nu }{160}+\frac{543 \nu ^2}{32}
\nonumber\\
&&
+\frac{4609 \nu ^3}{80}\bba) (p.n)^6
\bba] u^2
+ \bba[
        \bba(
                \frac{45}{128}-\frac{95 \nu }{32}+\frac{475 \nu ^2}{64}-\frac{267 \nu ^3}{64}-\frac{35 \nu ^4}{128}\bba) p^8
        +\bba(
                \frac{5 \nu }{32}-\frac{29 \nu ^2}{32}
\nonumber\\
&& +\frac{11 \nu ^3}{32}
-\frac{5 \nu ^4}{32}\bba) p^6 (p.n)^2
       +\big(
                -\frac{9 \nu ^2}{64}+\frac{33 \nu ^3}{64}-\frac{9 \nu ^4}{64}\big) p^4 (p.n)^4
        -\frac{5}{32} (-2+\nu ) \nu ^3 p^2 (p.n)^6
\nonumber\\
&&        -\frac{35}{128} \nu ^4 (p.n)^8
\Biggr] u
+\frac{1}{\varepsilon} \bbb [
\frac{2}{15} \nu  (-1+66 \nu ) u^5
+\bbb(
        -\frac{1}{180} \nu  (-1215+827 \nu ) p^2
 +\frac{1}{180} \nu  (-3354
\nonumber\\ &&
+10685 \nu ) (p.n)^2
\bbb)
 u^4
+ \bbb(
        -\frac{1}{180} \nu  (-1425+757 \nu ) p^4
        -\frac{11}{60} \nu  (195
+133 \nu ) p^2 (p.n)^2
\nonumber\\ &&
            +\frac{5}{3} \nu  (12+37 \nu ) (p.n)^4
\bbb) u^3 \bbb]
+ \ln \left(
        \frac{r}{r_0}
\right)
\bbb[\frac{4}{3} \nu  (-1+66 \nu ) u^5
+\bbb(
                -\frac{2}{45} \nu  (-1215+827 \nu ) p^2
\nonumber\\ &&    
                +\frac{2}{45} \nu  (-3354+10685 \nu ) (p.n)^2
        \bbb) u^4
        + \bbb(
                -\frac{1}{30} \nu  (-1425+757 \nu ) p^4
                -\frac{11}{10} \nu  (195
\nonumber\\ &&    
+133 \nu ) p^2 (p.n)^2
               +10 \nu  (12+37 \nu ) (p.n)^4
        \bbb) u^3
\bbb ]
\bbb \},
\end{eqnarray}
%-----
\begin{eqnarray}
\hat H_{\mathrm{4PN}}^{\mathrm{tail}} &=& \nu \bbb\{
\frac{1}{\varepsilon}\bbb [ 
u^4 \bba(
        -\frac{16 \nu  p^2}{5}
        +\frac{44}{15} \nu  (p.n)^2
\bba)
\bbb ]
+ \ln \left(
        \frac{r}{r_0}\right) u^4 \bba(
        -\frac{64 \nu  p^2}{5}
        +\frac{176}{15} \nu  (p.n)^2
\bba) \nonumber\\
&& + \bba[
        \frac{784 \nu  p^2}{75}
        -\frac{264}{25} \nu  (p.n)^2
\bba] u^4 + {\cal I}^{\mathrm{tail}}
\bbb\},
%----
\\
  {\cal I}^{\mathrm{tail}}&=&  
  - \frac{G_N^2 }{5 \nu}
 \dddot I_{ij}(t) \int_0^{+\infty} d\tau  \ln \left( \frac{ \tau }{\tau_0} \right ) \bigg ( \ddddot {I_{ij}} (t- 
\tau) - \ddddot {I_{ij}} (t+ \tau) \bigg )
\\
\text{and} & & \nonumber\\
  {\cal I}^{\mathrm{tail}}_{\rm circ}&=&  
 \frac{2 G_N^2   }{5\nu}  \left(\dddot
    I_{ij}(t) \right)^2 \big ( \gamma_E +  \ln{(2 \Omega \tau_0})
    \big)   \,,
\end{eqnarray}
%--------------------------------------------------------------------------------------------------------
%--------------------------------------------------------------------------------------------------------
with $p.n = p.r/r$, $r = |\vec{r}|$ and $\left(\dddot I_{ij}(t) \right)^2$ is given in Eq.~(\ref{eq:PM17}). 

The terms of first order 
in $\nu$ can be obtained in the Schwarzschild test-particle limit in harmonic coordinates \cite{WEINB,SCHMUTZER1}
%--------------------------------------------------------------------------------------------------------
\begin{eqnarray}
\hat{H}_{\rm TP} &=& 
\nu \Biggl\{\sqrt{\frac{1 - u}{1 + u}} \sqrt{1 + \frac{p^2 - (p.n)^2 u^2}{(1 + u)^2}}
-1 \Biggr\}
\nonumber\\ &=&
\nu \Biggl\{\Biggl[\frac{1}{2} p^2 - \frac{1}{8} (p^2)^2 + \frac{1}{16} (p^2)^3 - \frac{5}{128} (p^2)^4
+ \frac{7}{256} (p^2)^5 
\Biggr]
+ u \Biggl[-1 -\frac{3}{2} p^2 + \frac{5}{8} (p^2)^2  
\nonumber\\ &&
- \frac{7}{16} (p^2)^3 
+ \frac{45}{128} (p^2)^4 \Biggr]
+ u^2 \Biggl[
\frac{1}{2} 
+ \left(\frac{11}{4} p^2 - \frac{1}{2} (p.n)^2\right)
+ \left(-\frac{29}{16} (p^2)^2 + \frac{1}{4} p^2 (p.n)^2 \right) 
\nonumber\\ &&
+ \left(\frac{55}{32} (p^2)^3 - \frac{3}{16} (p^2)^2 (p.n)^2 \right)
\Biggr]
+ u^3 \Biggl[
-\frac{1}{2} + \left( -\frac{17}{4} p^2 +\frac{3}{2} (p.n)^2\right)
\nonumber\\ &&
+ \left(\frac{65}{16} (p^2)^2 - \frac{5}{4} p^2 (p.n)^2 \right)
\Biggr] 
+ u^4 \Biggl[ 
\left(\frac{95}{16} p^2 - \frac{11}{4} (p.n)^2\right)
\Biggr]
- \frac{3}{8} u^5 + O(5PN)
\Biggr\}~.
\end{eqnarray}
%--------------------------------------------------------------------------------------------------------
Both the 3PN and 4PN (effective) Hamiltonian contain pole-terms in the dimensional parameter $\ep$ and logarithmic 
contributions. This is not unusual since these quantities are not observables.\footnote{In a very vague sense these
Hamiltonians can be compared to the bare Hamiltonians in quantum field theory which in many cases are singular
in all their pieces, except the differential operators, but see Section~\ref{sec:24}.}
%--------------------------------------------------------------------------------------------------------
\subsection{The cancellation of the pole contributions}
\label{sec:24}
%--------------------------------------------------------------------------------------------------------

\vspace*{1mm}
\noindent
Using the effective field theory approach of Einstein--Hilbert gravity in the post--Newtonian orders,
the harmonic coordinates belong to a class in which pole terms already occur at the level of 3PN. 
During the whole calculation one works in $3-2\ep$ Euclidean space dimensions, starting at the Newtonian level.
In various places one has to expand up to $O(\ep)$ because certain terms appear multiplicatively together
with pole terms. Because of the possible diagrammatic structures, the 6PN terms can potentially contain double poles,
the 9PN terms triple poles etc., which is also indicated by the structure of the canonical transformations
given e.g. in (\ref{eq:POLfr4}).

On the other hand, ADM coordinates \cite{Damour:2014jta} and EOB coordinates \cite{Bini:2019nra} are pole--free 
at 3PN.~\footnote{At 4PN dimensional regularization was used for the local part of the ADM Hamiltonian in 
\cite{Jaranowski:2015lha}. The pole terms were absorbed into a total time derivative. The remaining contribution 
has been calculated in three dimensions. This actually implies a canonical transformation from ADM into another 
frame, cf.~\cite{Damour:1990jh}.} Since at 3PN a canonical 
transformation exists, cf. Section~\ref{sec:25},
transforming to pole--free coordinates, no observable will exhibit poles, and, at 3PN, associated to that, no
logarithmic terms of the kind $\ln(r/r_0)$. 

At 4PN for the complete Hamiltonian, including the tail terms, the poles arrange such, that they can be transformed away 
by a canonical transformation ending up with a pole--free
Hamiltonian. However, logarithmic contributions remain. The latter ones are free of the dimensional mass scale $\mu_1$, 
cf.~Section~\ref{sec:22}. Therefore, the potential renormalization group equation 
\cite{Symanzik:1970rt,Callan:1970yg} of the bare Hamiltonian is trivial and no renormalization takes place;
both the interacting point masses remain constant w.r.t. their rest mass and Newton's constant remains a constant, 
cf. also \cite{Alvey:2019ctk}.
%--------------------------------------------------------------------------------------------------------
\subsection{Hamiltonians in Pole--free Coordinates}
\label{sec:25}
%--------------------------------------------------------------------------------------------------------

\vspace*{1mm}
\noindent
Although first order bare Lagrangians and Hamiltonians of the effective field approach to classical gravity 
are allowed to have poles in the dimensional parameter $\ep$, there exist classes of coordinates which are 
free of poles. This is well--known at 3PN and examples are the ADM and EOB coordinates. For convenience, we 
provide a canonical transformation from harmonic coordinates to this class in the following. These Hamiltonians 
have no real preference against Hamiltonians with poles, as e.g. obtained in the harmonic gauge, since all
observables are pole free. In conjunction with the poles also associated logarithmic contributions emerge. These
related logarithms are eliminated in the same manner, while others, stemming from the tail terms, may remain.

Canonical transformations 
form a group, cf. e.g.~\cite{MITTELSTAEDT}, and one may therefore also consider their products for some purpose.
Let us consider the canonical transformation of a Hamiltonian $H$ to a Hamiltonian $H'$.
Given the differential operator $D_g$
%--------------------------------------------------------------------------------------------------------
\begin{eqnarray}
D_g = \sum_{k=1}^{2l} \beta_k \frac{\partial }{\partial y_k}.
\end{eqnarray}
%--------------------------------------------------------------------------------------------------------
By using $y_k = q_k$ for $1\le k \le l$ and $y_k = p_k$ for  $l+1 \le k \le 2l$, one can write the canonical transformation 
induced by the generator $g$ as Lie--series~\cite{GROEBNER,MITTELSTAEDT}\footnote{Lie series were also applied in celestial
mechanics in \cite{VINTI}.} in the following form,   
%--------------------------------------------------------------------------------------------------------
\begin{eqnarray}
y'_l = \sum_{n = 0}^\infty \frac{1}{n!}(D_g)^n y_l
\end{eqnarray}
%--------------------------------------------------------------------------------------------------------
or 
%--------------------------------------------------------------------------------------------------------
\begin{eqnarray}
y'_l = {\rm exp}[D_g] y_l,
\end{eqnarray}
%--------------------------------------------------------------------------------------------------------
with the linear differential operator 
%--------------------------------------------------------------------------------------------------------
\begin{eqnarray}
D_g = 
\sum_{n=1}^k \left[\frac{\partial }{\partial y_{n+k}}\frac{\partial g}{\partial y_n} - 
\frac{\partial }{\partial y_n}\frac{\partial g}{\partial y_{n+k}}\right].
\end{eqnarray}
%--------------------------------------------------------------------------------------------------------
By the aid of the fundamental relations 
%--------------------------------------------------------------------------------------------------------
\begin{eqnarray}
{\rm exp}[D_g](A + B) &=& {\rm exp}[D_g]A +  {\rm exp}[D_g]B
\\
{\rm exp}[D_g](AB)    &=& {\rm exp}[D_g](A)~{\rm exp}[D_g](B)
\end{eqnarray}
%--------------------------------------------------------------------------------------------------------
one obtains
%--------------------------------------------------------------------------------------------------------
\begin{eqnarray}
H'(y) = {\rm exp}[D_g] H(y) = H(y').
\end{eqnarray}
%--------------------------------------------------------------------------------------------------------
The canonical transformations are associated to symplectic groups \cite{JACOBI1,KILLING,CARTAN}.

The generating functions for the canonical transformation of $\hat{H}_{\rm harm}$ to the associated pole--free 
Hamiltonian $\hat{H}_{\rm harm}^{\rm pole free} $are given by
%--------------------------------------------------------------------------------------------------------
\begin{eqnarray}
\label{eq:POLfr4}
  G_{\rm 3L} &=& \bbb(
        -\frac{17 \nu }{6 \varepsilon }
        -17 \nu  \ln \left(
                \frac{r}{r_0}\right)
\bbb) u^2 p.n, 
\\
%--------------------------
  G_{\rm 4L} &=&  \bbb[
        (p.n)^3 \bba(
                -\frac{4 \nu }{\varepsilon }
                -\frac{37 \nu ^2}{3 \varepsilon }
                -24 \nu  \ln \big(
                        \frac{r}{r_0}\big)
                -74 \nu ^2 \ln \left(
                        \frac{r}{r_0}\right)
        \bba)
        +p^2 p.n \bba(
                \frac{13 \nu }{2 \varepsilon }
                +\frac{2 \nu ^2}{45 \varepsilon }
\nonumber\\ &&
                +39 \nu  \ln \left(
                        \frac{r}{r_0}\right)
                +\frac{4}{15} \nu ^2 \ln \left(
                        \frac{r}{r_0}\right)
        \bba)
\bbb] u^2
+ p.n \bbb[
        -\frac{27 \nu }{10 \varepsilon }
        -\frac{44 \nu ^2}{5 \varepsilon}
        -\frac{108}{5} \nu  \ln \left(
                \frac{r}{r_0}\right)
\nonumber\\ &&
        -\frac{352}{5} \nu ^2 \ln \left(
                \frac{r}{r_0}\right)
\bbb] u^3.
\end{eqnarray}
%--------------------------------------------------------------------------------------------------------
The transformations read
%--------------------------------------------------------------------------------------------------------
\begin{eqnarray}
\hat H_{\mathrm{3PN,harm}}^{\mathrm{polefree}} &=& \hat H_{\mathrm{3PN,harm}} + \{H_{\rm N},G_{\rm 3L}\}
\\
\label{eq:H4pfcan}
\hat H_{\mathrm{4PN,harm}}^{\mathrm{polefree}} &=& \hat H_{\mathrm{4PN,harm}} + \{H_{\rm 1PN,harm},G_{\rm 3L}\}
+\{H_{\rm N},G_{\rm 4L}\},
\end{eqnarray}
%--------------------------------------------------------------------------------------------------------
with the Poisson brackets
%--------------------------------------------------------------------------------------------------------
\begin{eqnarray}
\left\{A,B\right\} = \sum_{i=1}^N 
  \left(\frac{\partial A}{\partial r_i}
  \frac{\partial B}{\partial p_i}
- \frac{\partial A}{\partial p_i}
  \frac{\partial B}{\partial r_i}\right).
\end{eqnarray}
%--------------------------------------------------------------------------------------------------------

One obtains 
%--------------------------------------------------------------------------------------------------------
\begin{eqnarray}
\hat H_{\mathrm{3PN,harm}}^{\mathrm{polefree}} &=&
\nu \Biggl\{  \bba(
        -\frac{5}{128}+\frac{35 \nu }{128}-\frac{35 \nu \
^2}{64}+\frac{35 \nu ^3}{128}\bba) p^8
+\frac{1}{576} \big(
        216-5336 \nu +135 \nu  \pi ^2\big) u^4
\nonumber\\
&&+ \bbb[
        p^2 \bba(
                -\frac{17}{4}+\frac{643 \nu }{72}-\frac{3 \nu ^2}{2}-\
\frac{7 \nu  \pi ^2}{8}\bba)
        +\bba(
                \frac{3}{2}-\frac{809 \nu }{12}+\frac{49 \nu \
^2}{4}+\frac{21 \nu  \pi ^2}{8}\bba) 
\nonumber\\
&&
\times (p.n)^2
\bbb] u^3
+ \bbb[
        \frac{1}{16} (-1+8 \nu ) (29+12 \nu ) p^4
        +\frac{1}{4} \big(
                1-36 \nu -36 \nu ^2\big) p^2 (p.n)^2
\nonumber\\
&&
 +\frac{1}{3} \nu  (7+69 \nu ) (p.n)^4
\bbb] u^2
+  \bbb[
        \bba(
                -\frac{7}{16}+\frac{45 \nu }{16}-\frac{31 \nu ^2}{8}-\
\frac{5 \nu ^3}{16}\bba) p^6
        +\bba(
                -\frac{3 \nu }{16}
\nonumber\\
&& 
+\frac{11 \nu ^2}{16}-\frac{3 \nu \
^3}{16}\bba) p^4 (p.n)^2
       -\frac{3}{16} (-1+\nu ) \nu ^2 p^2 (p.n)^4
        -\frac{5}{16} \nu ^3 (p.n)^6
\bbb] u \Biggr\},
\\
%----
\hat H_{\mathrm{4PN,harm}}^{\mathrm{polefree}} &=&
\nu \Biggl\{ \bbb(
        \frac{7}{256}-\frac{63 \nu }{256}+\frac{189 \nu \
^2}{256}-\frac{105 \nu ^3}{128}+\frac{63 \nu ^4}{256}\bbb) p^{10}
\nonumber\\
&&+\bbb(
        -\frac{3}{8}-\frac{42571 \nu }{400}+\frac{43907 \nu ^2}{400}+\
\frac{79 \nu  \pi ^2}{64}-\frac{59 \nu ^2 \pi ^2}{8}\bbb) u^5
\nonumber\\
&&
+  \bbb[
        p^2 \bba(
                \frac{95}{16}+\frac{509593 \nu }{14400}-\frac{1173683 \
\nu ^2}{7200}+\frac{643 \nu  \pi ^2}{128}+\frac{1419 \nu ^2 \pi \
^2}{128}\bba)
\nonumber\\
&&        +\bba(
                -\frac{11}{4}+\frac{3733 \nu }{320}+\frac{1880651 \nu \
^2}{7200}-\frac{4429 \nu  \pi ^2}{192}-\frac{4033 \nu ^2 \pi ^2}{128}\
\bba) (p.n)^2
\bbb] u^4
\nonumber\\
&&
+  \bbb[
        p^4 \bba(
                \frac{65}{16}-\frac{94439 \nu }{800}+\frac{319789 \nu \
^2}{14400}+\frac{205 \nu ^3}{32}+\frac{1091 \nu  \pi \
^2}{1024}-\frac{217 \nu ^2 \pi ^2}{64}\bba)
\nonumber\\
&&        +(p.n)^4 \bba(
                -\frac{5927 \nu }{32}-\frac{176689 \nu \
^2}{320}-\frac{333 \nu ^3}{32}+\frac{4395 \nu  \pi \
^2}{1024}+\frac{345 \nu ^2 \pi ^2}{128}\bba)
\nonumber\\
&&        +p^2 (p.n)^2 \bba(
                -\frac{5}{4}+\frac{256477 \nu }{800}+\frac{197453 \nu \
^2}{1200}+\frac{11 \nu ^3}{16}-\frac{2955 \nu  \pi \
^2}{512}+\frac{1095 \nu ^2 \pi ^2}{128}\bba)
\bbb] u^3
\nonumber\\
&&+  \bbb[
        \bba(
                \frac{55}{32}-\frac{667 \nu }{64}+\frac{1217 \nu \
^2}{64}-\frac{89 \nu ^3}{64}\bba) p^6
        +\bba(
                -\frac{3}{16}-\frac{99 \nu }{16}+\frac{733 \nu \
^2}{16}
\nonumber\\
&& 
+\frac{3189 \nu ^3}{64}\bba) p^4 (p.n)^2
       +\bba(
                -\frac{79 \nu }{192}-\frac{4737 \nu \
^2}{64}-\frac{7511 \nu ^3}{96}\bba) p^2 (p.n)^4
        +\bba(
                \frac{487 \nu }{160}+\frac{543 \nu ^2}{32}
\nonumber\\
&&
+\frac{4609 \
\nu ^3}{80}\bba) (p.n)^6
\bbb] u^2
+  \bbb[
        \bba(
                \frac{45}{128}-\frac{95 \nu }{32}+\frac{475 \nu \
^2}{64}-\frac{267 \nu ^3}{64}-\frac{35 \nu ^4}{128}\bba) p^8
        +\bba(
                \frac{5 \nu }{32}
\nonumber\\
&& 
-\frac{29 \nu ^2}{32}+\frac{11 \nu \
^3}{32}-\frac{5 \nu ^4}{32}\bba) p^6 (p.n)^2
       +\bba(
                -\frac{9 \nu ^2}{64}+\frac{33 \nu ^3}{64}-\frac{9 \nu \
^4}{64}\bba) p^4 (p.n)^4
\nonumber\\ &&        
-\frac{5}{32} (-2+\nu ) \nu ^3 p^2 (p.n)^6
        -\frac{35}{128} \nu ^4 (p.n)^8
\bbb] u + {\cal J}^{\mathrm{tail}} \Biggr\},
\\
\text{with} & & \nonumber\\ 
{\cal J}^{\mathrm{tail}} &=& {\cal I}_{\rm tail} + \frac{2}{5\nu} G_N^2  (\dddot I_{ij}(t))^2 \ln 
\left(\frac{r}{r_0}\right).
\label{eq:JTAIL}
\end{eqnarray}
%--------------------------------------------------------------------------------------------------------
Note that here the contributions to the Hamiltonian are used in $D$ dimensions. Unlike the case in (\ref{eq:ADM4}) 
a double commutator does not contribute to (\ref{eq:H4pfcan}). It emerges for the first time at 6PN.
Since canonical transformations leave the dynamics of a system invariant \cite{HAMJAC}, one obtains that 
all observables calculated using (\ref{eq:H0}--\ref{eq:H4}) lead to the same results.
Here the canonical transformations may contain also the regularization parameter $\ep \neq 0$,
which would formally imply singularities in the limit $\ep \rightarrow 0$ in the Hamiltonian. However, this is
not relevant since all observables do not depend on this parameter. It only remains a necessary asset by applying
dimensional regularization for the calculation. 
%-------------------------------------------------------------------------------------------------------- 
\section{ADM coordinates} 
\label{sec:3} 
%-------------------------------------------------------------------------------------------------------- 

\vspace*{1mm} 
\noindent 
We notice that $\hat{H}_N$ and $\hat{H}_{1PN}$ in harmonic coordinates are the same as in ADM coordinates 
\cite{Damour:2014jta}. 
Now we construct the canonical transformation from our results to those of Ref.~\cite{Damour:2014jta}
using the same variables at both sides. Both Hamiltonians are then equivalent and lead to
the same observables, which are also frame-independent \cite{POINEIN}. 

The generating functions for the canonical transformation from $\hat{H}_{\rm harm}$ to $\hat{H}_{\rm ADM}$
are given by
%--------------------------------------------------------------------------------------------------------
\begin{eqnarray}
  G_{2} &=& p.n \Biggl[-\frac{1}{4} \nu  p^2 
+\frac{1}{4} (1-2 \nu )  u \Biggr],
\\
%----
  G_{3} &=& p.n \Biggl[\bba[
        -\frac{1}{16} \nu  (-57+10 \nu ) p^2 
        -\frac{1}{48} \nu  (23+224 \nu ) (p.n)^2
\bba] u 
\nonumber\\
&&-\frac{1}{16} \nu  (-1+3 \nu ) p^4 
-\frac{1}{576} \nu  \big(
        -10280+513 \pi ^2\big) u^2 	\Biggr]
 + G_{\rm 3L},\\
%----
G_{4} &=& 
p.n \Biggl\{\bba(
        \frac{1}{32}+\frac{388531 \nu }{7200}-\frac{492943 \nu \
^2}{3600}+\frac{4061 \nu  \pi ^2}{1024}+\frac{30059 \nu ^2 \pi \
^2}{3072}\bba) u^3 
\nonumber\\
&&
 + \bbb[
        p^2 \bba(
                -\frac{2856641 \nu }{57600}+\frac{506003 \nu \
^2}{19200}+\frac{2331 \nu  \pi ^2}{8192}-\frac{52155 \nu ^2 \pi \
^2}{16384}\bba)
\nonumber\\
&&        +\bba(
                \frac{633481 \nu }{19200}+\frac{1930769 \nu \
^2}{19200}-\frac{6957 \nu  \pi ^2}{8192}-\frac{16563 \nu ^2 \pi \
^2}{16384}\bba) (p.n)^2
\bbb] u^2
\nonumber\\
&&
+  \bbb[
        \bba(
                \frac{173 \nu }{32}-\frac{2789 \nu \
^2}{256}-\frac{1133 \nu ^3}{128}\bba) p^4 
        +\bba(
                -\frac{139 \nu }{96}+\frac{5545 \nu \
^2}{384}+\frac{3113 \nu ^3}{192}\bba) p^2 (p.n)^2
\nonumber\\
&&        +\bba(
                -\frac{59 \nu }{480}-\frac{3323 \nu \
^2}{768}-\frac{15559 \nu ^3}{1920}\bba) (p.n)^4
\bbb] u
+\bba(
        -\frac{\nu }{32}+\frac{5 \nu ^2}{32}+\frac{137 \nu \
^3}{256}\bba) p^6 
\nonumber\\
&&-\frac{59}{768} \nu ^3 p^4 (p.n)^2
-\frac{13}{256} \nu ^3 p^2 (p.n)^4
+\frac{5}{256} \nu ^3 (p.n)^6 \Biggr\}
+ G_{\rm 4L}.
\end{eqnarray}
%--------------------------------------------------------------------------------------------------------
The Hamiltonians starting from 2PN are related by
%--------------------------------------------------------------------------------------------------------
\begin{eqnarray}
\hat{H}_{\rm ADM}^{(2)} &=& \hat{H}_{\rm harm}^{(2)}  + \{\hat{H}_{\rm N},G_2\},
\\ 
%---
\label{eq:ADM3}
\hat{H}_{\rm ADM}^{(3)} &=& \hat{H}_{\rm harm}^{(3)}  
+ \{\hat{H}_{\rm N},G_3\}
+ \{\hat{H}_{\rm harm}^{(1)},G_2\},
\\ 
%---
\label{eq:ADM4}
\hat{H}_{\rm ADM}^{(4)} &=& \hat{H}_{\rm harm}^{(4)}
+ \{\hat{H}_{\rm N},G_4\}
+ \{\hat{H}_{\rm harm}^{(1)},G_3\}
+ \{\hat{H}_{\rm harm}^{(2)},G_2\}
+ \{\{\hat{H}_{\rm N},G_2\},G_2\}.
\end{eqnarray}
%--------------------------------------------------------------------------------------------------------
%-------------------------------------------------------------------------------------------------------- 
\section{Effective one body coordinates} 
\label{sec:4} 
%-------------------------------------------------------------------------------------------------------- 

\vspace*{1mm} 
\noindent 
We will now consider the link to the Hamiltonian of effective one body (EOB) theory 
\cite{Buonanno:1998gg,Buonanno:2000ef}, which is widely used. Here we consider the effective Hamiltonian 
\cite{Buonanno:1998gg}, 
Eq.~(6.5), 
and transform from ADM to EOB coordinates. We define
%-------------------------------------------------------------------------------------------------------- 
\begin{eqnarray} 
\label{eq:EOBXH} 
\hat{H}_{\rm EOB}^{\rm eff} = \frac{H_{\rm EOB}}{\mu c^2}. 
\end{eqnarray} 
%--------------------------------------------------------------------------------------------------------
Note that in \cite{Damour:2015isa}, Eq.~(6.1), $\alpha_3 = \alpha_4 = 0$ and 
%-------------------------------------------------------------------------------------------------------- 
\begin{eqnarray} 
\label{eq:EOB1H} 
\hat{H}_{\rm EOB}^{\rm eff} 
= \sqrt{ A [1 + A \bar{D} (p.n)^2 + (p^2 - (p.n)^2) + Q]} = 1 + 
\frac{1}{c^2 \nu}\hat{H}_{\rm ADM} \left[1 + \frac{1}{c^2}\frac{1}{2} \hat{H}_{\rm ADM} \right], 
\end{eqnarray} 
%--------------------------------------------------------------------------------------------------------
with the functions $A, \bar{D}$ and $Q$ given in \cite{Bini:2019nra}, 
cf.~\cite{Buonanno:1998gg}, Eq.~(6.10). Note that one also has  to rescale 
%-------------------------------------------------------------------------------------------------------- 
\begin{eqnarray} 
p^2 \rightarrow p^2/c^2,~~~(p.n)^2 \rightarrow (p.n)^2/c^2,~~~u \rightarrow u/c^2,~~~r \rightarrow r c^2  
\label{eq:EOB2H} 
\end{eqnarray} 
%--------------------------------------------------------------------------------------------------------
in \cite{Bini:2019nra} in the root-term in (\ref{eq:EOB1H}), to be able to expand the contributions to 
(\ref{eq:EOB1H}) to compare the respective contributions. Here the coordinates in the 
r.h.s. are those of ADM and in the l.h.s. of EOB. In the following we will first consider the local contributions. 
Furthermore, it is convenient to consider the canonical transformation form ADM to the EOB coordinates for the
quantity 
%--------------------------------------------------------------------------------------------------------
\begin{eqnarray}
\label{eq:EOB3}
T_{\rm EOB} \equiv \frac{1}{2 c^4}(\hat{H}_{\rm EOB}^{\rm eff})^2 
&=& \frac{1}{2} + \sum_{k=0}^4 \frac{1}{c^{2(k+1)}} T_{\rm EOB}^{(k)}
= \frac{1}{2} + \sum_{k=0}^4 \frac{1}{c^{2(k+1)}} T_{\rm ADM}^{(k)},
\nonumber\\ &=& 
\frac{1}{2} +  \frac{1}{c^2 \nu}\hat{H}_{\rm ADM} + 
\frac{1}{2c^4 \nu^2}(1+\nu)  \hat{H}_{\rm ADM}^2 + \frac{1}{2c^6 \nu^2} \hat{H}_{\rm ADM}^3 + \frac{1}{8c^8 \nu^2} 
\hat{H}_{\rm ADM}^4,
\nonumber\\
\end{eqnarray}
%--------------------------------------------------------------------------------------------------------
with 
%--------------------------------------------------------------------------------------------------------
\begin{eqnarray}
\hat{H}_{\rm ADM} &=& 
\hat{H}_{\rm N} +
\frac{1}{c^2} \hat{H}_{\rm ADM,1PN} +
\frac{1}{c^4}\hat{H}_{\rm ADM,2PN} +
\frac{1}{c^6}\hat{H}_{\rm ADM,3PN} +
\frac{1}{c^8}\hat{H}_{\rm ADM,4PN}. 
\end{eqnarray}
%--------------------------------------------------------------------------------------------------------
The different post--Newtonian contributions in both sides of (\ref{eq:EOB3}) are now labeled by powers of $1/c^2$.
We determine the canonical transformation mapping the expansion coefficients of 
$T_{\rm ADM}^{(k)}$ to $T_{\rm EOB}^{(k)}$.\footnote{In \cite{Buonanno:1998gg} a corresponding canonical transformation 
has been constructed between $\hat{H}_{\rm ADM}^{(k)}$ to $T_{\rm EOB}^{(k)}$ up to the 2nd post--Newtonian order, which 
is equivalent to the present approach at this level.} 
We write the canonical transformations using the respective terms
of the associated Lie--series \cite{GROEBNER,MITTELSTAEDT}.

At the Newtonian order $T_{\rm ADM}^{(0)} = \hat{H}_{\rm N}$ and $T_{\rm EOB}^{(0)}$ are the same. In the 
post--Newtonian orders
one has
%--------------------------------------------------------------------------------------------------------
\begin{eqnarray}
T_{\rm EOB}^{(1)} &=& T_{\rm ADM}^{(1)}  + \{T_{\rm ADM}^{(0)},g_1\},
\\ 
%---
T_{\rm EOB}^{(2)} &=& T_{\rm ADM}^{(2)}  
+ \{T_{\rm ADM}^{(1)},g_1\}
+ \{T_{\rm ADM}^{(0)},g_2\}
+ \frac{1}{2}\{\{T_{\rm ADM}^{(0)},g_1\},g_1\},
\\ 
%---
T_{\rm EOB}^{(3)} &=& T_{\rm ADM}^{(3)}  
+ \{T_{\rm ADM}^{(2)},g_1\}
+ \{T_{\rm ADM}^{(1)},g_2\}
+ \{T_{\rm ADM}^{(0)},g_3\}
+ \frac{1}{2}\Biggl[\{\{T_{\rm ADM}^{(1)},g_1\},g_1\}
\nonumber\\ &&
+ \{\{T_{\rm ADM}^{(0)},g_2\},g_1\}
+ \{\{T_{\rm ADM}^{(0)},g_1\},g_2\}\Biggr]
+ \frac{1}{6}\{\{\{T_{\rm ADM}^{(0)},g_1\},g_1\},g_1\},
\\ 
%---
T_{\rm EOB}^{(4)} &=& T_{\rm ADM}^{(4)}  
+ \{T_{\rm ADM}^{(3)},g_1\}
+ \{T_{\rm ADM}^{(2)},g_2\}
+ \{T_{\rm ADM}^{(1)},g_3\}
+ \{T_{\rm ADM}^{(0)},g_4\}
\nonumber\\ &&
+ \frac{1}{2}\Biggl[\{\{T_{\rm ADM}^{(2)},g_1\},g_1\}
+ \{\{T_{\rm ADM}^{(1)},g_2\},g_1\}
+ \{\{T_{\rm ADM}^{(1)},g_1\},g_2\} 
+ \{\{T_{\rm ADM}^{(0)},g_2\},g_2\}
\nonumber\\ &&
+ \{\{T_{\rm ADM}^{(0)},g_3\},g_1\}
+ \{\{T_{\rm ADM}^{(0)},g_1\},g_3\} 
\Biggr]
+ \frac{1}{6}\Biggl[
  \{\{\{T_{\rm ADM}^{(1)},g_1\},g_1\},g_1\}
\nonumber\\ &&
+ \{\{\{T_{\rm ADM}^{(0)},g_2\},g_1\},g_1\}
+ \{\{\{T_{\rm ADM}^{(0)},g_1\},g_2\},g_1\}
+ \{\{\{T_{\rm ADM}^{(0)},g_1\},g_1\},g_2\}
\Biggr]
\nonumber\\ &&
+ \frac{1}{24}
  \{\{\{\{T_{\rm ADM}^{(0)},g_1\},g_1\},g_1\},g_1\}.
\label{eq:EOB4PN}
%---
\end{eqnarray}
%--------------------------------------------------------------------------------------------------------
We rescale $r \rightarrow G_N M r$ and obtain 
the generating functions $\left. g_i\right|_{i=1}^4$ of the following canonical transformations 
%--------------------------------------------------------------------------------------------------------
\begin{eqnarray}
g_1(\vec{p},\vec{r}) &=& p.r \left\{\frac{\nu}{2} p^2 - \left[1+\frac{\nu}{2}\right]\frac{1}{r}\right\},
\\
%---
g_2(\vec{p},\vec{r}) &=& p.r \left\{
- \frac{\nu}{8} p^4
- \left[
   \frac{1}{4} (4 - \nu) \nu p^2 + \frac{3}{8} \nu (4 + \nu) (p.n)^2\right] \frac{1}{r}
- \frac{1}{4}(1 - 7 \nu + \nu^2) \frac{1}{r^2} 
\right\},
\\
%---
g_3(\vec{p},\vec{r}) &=& p.r \Biggl\{
        \frac{1}{16} (1-\nu ) \nu  p^6
%---
        +\Biggl[\frac{1}{96} \nu  \big(
                36-25 \nu -2 \nu ^2\big) p^4
        +\frac{1}{96} \nu  \big(
                24+16 \nu +\nu ^2\big) p^2 (p.n)^2
\nonumber\\ &&
        -\frac{1}{4} \nu ^2 (p.n)^4
        \Biggr] \frac{1}{r}
%---
        +\left[\frac{1}{48} \nu  \big(
                73-2 \nu +10 \nu ^2\big) p^2
        +\frac{1}{48} \nu  \big(
                87+28 \nu -6 \nu ^2\big) (p.n)^2
        \right]\frac{1}{r^2}
%---
\nonumber\\ &&
        -\frac{1}{192} \big(
                24
                -\nu  \big(
                        1396+3 \pi ^2\big)
                +12 \nu ^2
                +36 \nu ^3
        \big)\frac{1}{r^3}
\Biggr\},
%--------------
\\
g_4(\vec{p},\vec{r}) &=& 
p.r~\nu \Biggl\{
        -\frac{1}{384} \big(
                15-30 \nu +8 \nu ^2\big) p^8
%----
        + \Biggl\{-\frac{1}{768} \big(
                192-316 \nu +49 \nu ^2\big) p^6
\nonumber\\ &&
        - \Biggl[
                \frac{3}{16}-\frac{\nu }{48}-\frac{109 \nu ^2}{768}
        -\frac{\nu ^3}{96}\Biggr] p^4 (p.n)^2
        +\Biggl[
                \frac{\nu }{4}+\frac{13 \nu ^2}{256}-\frac{7 \nu^3}{128}\Biggr] p^2 (p.n)^4
\nonumber\\ &&
       -\frac{5}{256} \nu ^2 (1-2 \nu ) (p.n)^6
        \Biggr\}
        \frac{1}{r}
%----
        +\Biggl\{
         -\Biggl[
                \frac{101}{64}
                -\frac{3139}{768} \nu
                +\frac{97}{384} \nu^2
                +\frac{1}{24} \nu^3
         \Biggr] p^4
\nonumber\\ &&       
- \Biggl[ 
                \frac{17}{48}
                +\frac{243}{128} \nu
                -\frac{127}{192} \nu^2
                -\frac{1}{48} \nu^3
         \Biggr] p^2 (p.n)^2
       - \Biggl[
                \frac{593}{960}
                -\frac{16987}{3840} \nu
                +\frac{319}{640} \nu^2
                -\frac{1}{48} \nu^3 
         \Biggr] 
\nonumber\\ && \times    
(p.n)^4
\Biggr\}  
\frac{1}{r^2}
%----        
+\Biggl\{
        -\Biggl[
                 \frac{56163}{6400}
                +\frac{367553}{57600} \nu
                -\frac{67}{192} \nu^2
                -\frac{5}{24} \nu^3
                -\frac{(5626+18043 \nu ) \pi ^2}{16384}
        \Biggr] p^2
\nonumber\\ && 
       +\Biggl[
                \frac{88099}{19200}
                -\frac{265601}{19200} \nu
                +\frac{13}{64} \nu^2
                -\frac{1}{8} \nu^3
                -\frac{3 (50-2449 \nu ) \pi ^2}{16384}
        \Biggr] (p.n)^2
\Biggr\}
        \frac{1}{r^3}
\nonumber\\ &&
        -\Biggl\{        \frac{7}{96 \nu} 
-\frac{6409}{2400}
        +\frac{557}{360} \nu
        -\frac{3}{32} \nu^2
        +\frac{1}{6} \nu^3
        +\frac{(5886+3443 \nu ) \pi ^2}{3072}
        \Biggr\} \frac{1}{r^4} \Biggr\}.
%--------------
\end{eqnarray}
%--------------------------------------------------------------------------------------------------------

In this way we can directly compare to results obtained by using the effective one--body approach,
which have been worked out to the second post--Newtonian order in \cite{Buonanno:1998gg}, at third \cite{THIRD}
and fourth post--Newtonian order \cite{Damour:2015isa}, with first results at the fifth post--Newtonian order in 
\cite{Bini:2019nra}.

For convenience we display the Hamiltonian of EOB in closed form. Its principal structure has been given in 
\cite{Damour:2015isa}, Eq.~(2.12),
%--------------------------------------------------------------------------------------------------------
\begin{eqnarray}
\frac{1}{c^2}\hat{H}_{\rm EOB} &=& 
\sqrt{1+2\nu\left(\hat{H}^{\rm eff}_{\rm EOB}-1\right)}
\nonumber\\ &=&
1 + \frac{1}{c^2}\hat{H}_N 
%---
- \frac{\nu}{c^4}
\Biggl\{
  \frac{1}{8} (1+\nu) p^4
+ \left[\frac{1}{2} (1-\nu ) p^2 + (p.n)^2\right] \frac{1}{r}
+ \frac{1+\nu }{2 r^2}
\Biggr\}
\nonumber\\ &&
%-----
+ \frac{\nu}{c^6} \Biggl\{
\frac{1}{16} \big(
        1+\nu +\nu ^2\big) p^6
+ \left[\frac{1}{8} \big(
        1+\nu -3 \nu ^2\big) p^4
+\frac{1}{2} (1+\nu ) p^2 (p.n)^2\right] \frac{1}{r}
\nonumber\\ &&
- \left[\frac{1}{4} \big(
        1+\nu -3 \nu ^2\big) p^2
-(1+2 \nu ) (p.n)^2 \right] 
\frac{1}{r^2}
-\frac{1-\nu +\nu ^2}{2 r^3} 
\Biggr\} 
%-----
\nonumber\\ && 
- \frac{\nu}{c^8}\Biggl\{
 \frac{1}{128} \big(
        5+5 \nu +6 \nu ^2+5 \nu ^3\big) p^8
+\left[\frac{1}{16} \big(
        1+\nu -5 \nu ^3\big) p^6
+\frac{3}{8} \big(
        1+\nu +\nu ^2\big) p^4 (p.n)^2 \right]
\frac{1}{r}
\nonumber\\ &&
-\left[
\frac{1}{16} \big(
        1+\nu +6 \nu ^2-15 \nu ^3\big) p^4
-\frac{1}{2} (1+4 \nu ) p^2 (p.n)^2
-\frac{1}{2} (1-\nu ) (1-6 \nu ) (p.n)^4
\right]\frac{1}{r^2}
\nonumber\\ &&
+\left[\frac{1}{4} \big(
        1-\nu +2 \nu ^2-5 \nu ^3\big) p^2
-\frac{1}{2} \big(
        1+37 \nu -3 \nu ^2\big) (p.n)^2\right]
\frac{1}{r^3}
\nonumber\\ &&
+
\left[
        5
        -\frac{1}{24} \nu  \big(
                3080-123 \pi ^2\big)
        -2 \nu ^2
        +5 \nu ^3
\right]
\frac{1}{8 r^4} 
\Biggr\}
\nonumber\\ &&
%-----
+ \frac{\nu}{c^{10}} \Biggl\{
\frac{1}{256} \big(
        7+7 \nu +9 \nu ^2+10 \nu ^3+7 \nu ^4\big) p^{10}
%--
+\Biggl[
 \frac{1}{128} \big(
        5+5 \nu +3 \nu ^2-10 \nu ^3-35 \nu ^4\big) p^8
\nonumber\\ &&
+\frac{1}{16} \big(
        5+5 \nu +6 \nu ^2+5 \nu ^3\big) p^6 (p.n)^2
\Biggr] \frac{1}{r}
+\Biggl[\frac{1}{32} \big(
        -1-\nu -3 \nu ^2-10 \nu ^3+35 \nu ^4\big) p^6
%--
\nonumber\\ &&
-\frac{3}{8} (1+2 \nu ) \big(
        -1-2 \nu +\nu ^2\big) p^4 (p.n)^2
%--
+\frac{1}{4} \big(
        3-5 \nu +\nu ^2+6 \nu ^3\big) p^2 (p.n)^4
%--
\nonumber\\ &&
+\left[\frac{3}{10} \nu  \big(
        -3-9 \nu +10 \nu ^2\big) (p.n)^6
\right]
\frac{1}{r^2}
%--
+\Biggl[
\frac{1}{16} \big(
        1-\nu +\nu ^2+16 \nu ^3-35 \nu ^4\big) p^4
\nonumber\\ &&
+\frac{1}{4} \big(
        -1-37 \nu -36 \nu ^2+3 \nu ^3\big) p^2 (p.n)^2
+\frac{1}{2} \big(
        1+19 \nu -66 \nu ^2+4 \nu ^3\big) (p.n)^4
\Biggr]\frac{1}{r^3}
\nonumber\\ &&
+\Biggl[\frac{1}{384} \big(
        -120
        -336 \nu ^3
        +840 \nu ^4
        +\nu  \big(
                3080-123 \pi ^2\big)
        +\nu ^2 \big(
                -2888+123 \pi ^2\big)
\big) p^2
\nonumber\\ &&
+ \Biggl(\frac{1}{2} 
+  \left(\frac{611}{18} - \frac{25729}{3072}\pi^2\right) \nu
+ \left(-\frac{201}{2} + \frac{123}{32} \pi^2 \right) \nu^2
- \nu^3  \Biggr) (p.n)^2\Biggr] \frac{1}{r^4}
\nonumber\\ &&
- \left[
        7
        + \frac{\big(
                292096-24285 \pi ^2\big)}{1920} \nu
        +19 \nu ^2
        -2 \nu ^3
        +7 \nu ^4
\right] \frac{1}{8r^5} + {\cal J}_{\rm tail} \Biggr\},
\label{eq:HHeob}
\end{eqnarray} 
%--------------------------------------------------------------------------------------------------------
The local contribution to Eq.~(\ref{eq:HHeob}) has been derived from relations presented in Ref.~\cite{Bini:2019nra}.
The tail term ${\cal J}_{\rm tail}$ at 4PN is the same as in Eq.~(\ref{eq:JTAIL}).
%--------------------------------------------------------------------------------------------------------
%--------------------------------------------------------------------------------------------------------
\section{Observables at the fourth post-Newtonian level}
\label{sec:5}
%--------------------------------------------------------------------------------------------------------

\vspace*{1mm}
\noindent
In the following we calculate a series of observables up to 4PN starting from the Lagrangian in harmonic 
coordinates.\footnote{By removing the acceleration contributions already deviations to the original harmonic 
coordinates have been implied, cf.~\cite{Damour:1990jh}.} We present the energy of the innermost stable orbit $E(\Omega)$ 
as a function of the orbital frequency $\Omega$ and the angular momentum $J(\Omega)$ along circular orbits, since we would 
like to use the tail term in closed analytic form directly from harmonic coordinates. Otherwise one would have to
perform expansions in the excentricity \cite{Damour:2015isa,Damour:2016abl,EXC}.\footnote{For related results see also
\cite{Antonelli:2019ytb}.} We agree with the previous results given 
in \cite{Damour:2014jta}, derived by using different calculation methods. The following kinematic decomposition holds 
%--------------------------------------------------------------------------------------------------------
\begin{eqnarray}
p^2 = (p.n)^2 + \frac{J^2}{r^2}.
\end{eqnarray}
%--------------------------------------------------------------------------------------------------------

For circular orbits, $p.n = 0$, the energy $E$ is obtained by 
%--------------------------------------------------------------------------------------------------------
\begin{eqnarray}
E = {H}(r,J) - M.
\label{eq:EH}
\end{eqnarray}
%--------------------------------------------------------------------------------------------------------
The relation between the angular momentum $J$ and $r$ is found 
by 
%--------------------------------------------------------------------------------------------------------
\begin{eqnarray}
\frac{\partial H(r,J)}{\partial r}  = 0, 
\end{eqnarray}
%--------------------------------------------------------------------------------------------------------
cf.~\cite{Schafer:2018kuf}. 
One may express $E = E(J,\nu)$ using the variable  $x$ given in Eq.~(\ref{eq:x}) and obtains
%--------------------------------------------------------------------------------------------------------
\begin{eqnarray}
\label{eq:ENER}
\frac{E(x,\nu)}{\mu} &=& - \frac{1}{2} x \Biggl\{1 - \Biggl(\frac{3}{4} + \frac{\nu}{12}\Biggr) x
                  - \Biggl(\frac{27}{8} - \frac{19 \nu}{8} + \frac{\nu^2}{24} \Biggr)  x^2
                  - \Biggl[\frac{675}{64} -  \Biggl(\frac{34445}{576} - \frac{205 \pi^2}{96} \Biggr) \nu
\nonumber\\ &&                  
+ \frac{155}{96} \nu^2 + \frac{35}{5184} \nu^3\Biggr] x^3
                  + \Biggl[-\frac{3969}{128} + \Biggl(\frac{9038 \pi^2}{1536} - \frac{123671}{5760} + \frac{448}{15} [2 
                    \gamma_E + \ln(16 x)]\Biggr) \nu 
\nonumber\\ &&              
    + \Biggl(\frac{3157 \pi^2}{576} - \frac{498449}{3456} \Biggr) \nu^2 + \frac{301}{1728} \nu^3 + \frac{77}{31104} 
\nu^4 \Biggr] x^4 + O(x^5) \Biggr\}.
\end{eqnarray}
%--------------------------------------------------------------------------------------------------------
The innermost stable circular orbit is defined by the requirement $dE/dx = 0.$ 
The test particle solution reads \cite{Schafer:2018kuf}
%--------------------------------------------------------------------------------------------------------
\begin{eqnarray}
\frac{E(x)}{\mu} &=&  \frac{1-2x}{\sqrt{1-3x}} -1.
\label{eq:ENERTP }
\end{eqnarray}
%--------------------------------------------------------------------------------------------------------

It is convenient to normalize the angular momentum $J$ by $j = J/(G_N m_1 m_2)$, for which we obtain
%--------------------------------------------------------------------------------------------------------
\begin{eqnarray}
\label{eq:ANG}
\frac{1}{j^2(x,\nu)} &=& x \Biggl\{1 - \left(3 + \frac{\nu}{3}\right) x  + \frac{25}{4} \nu x^2 + \left(\frac{5269}{72} - 
\frac{41\pi^2}{12} - \frac{61 \nu}{12} + \frac{\nu^2}{81} \right) \nu x^3
+ \Biggl[\frac{18263 \pi^2}{768} 
\nonumber\\ &&
- \frac{1294339}{2880} + \frac{128}{3}[2\gamma_E + \ln(16 x)]+ \left(\frac{2747 \pi^2}{288}
- \frac{90985}{432}\right) \nu + \frac{181 \nu^2}{108} 
+ \frac{\nu^3}{243} \Biggr] \nu x^4  
\nonumber\\ &&
+ O(x^5) \Biggr\}.
%--------------------------------
\end{eqnarray}
%--------------------------------------------------------------------------------------------------------
%--------------------------------------------------------------------------------------------------------
\begin{figure}[H]
  \centering
  \hskip-0.8cm
  \includegraphics[width=.8\linewidth]{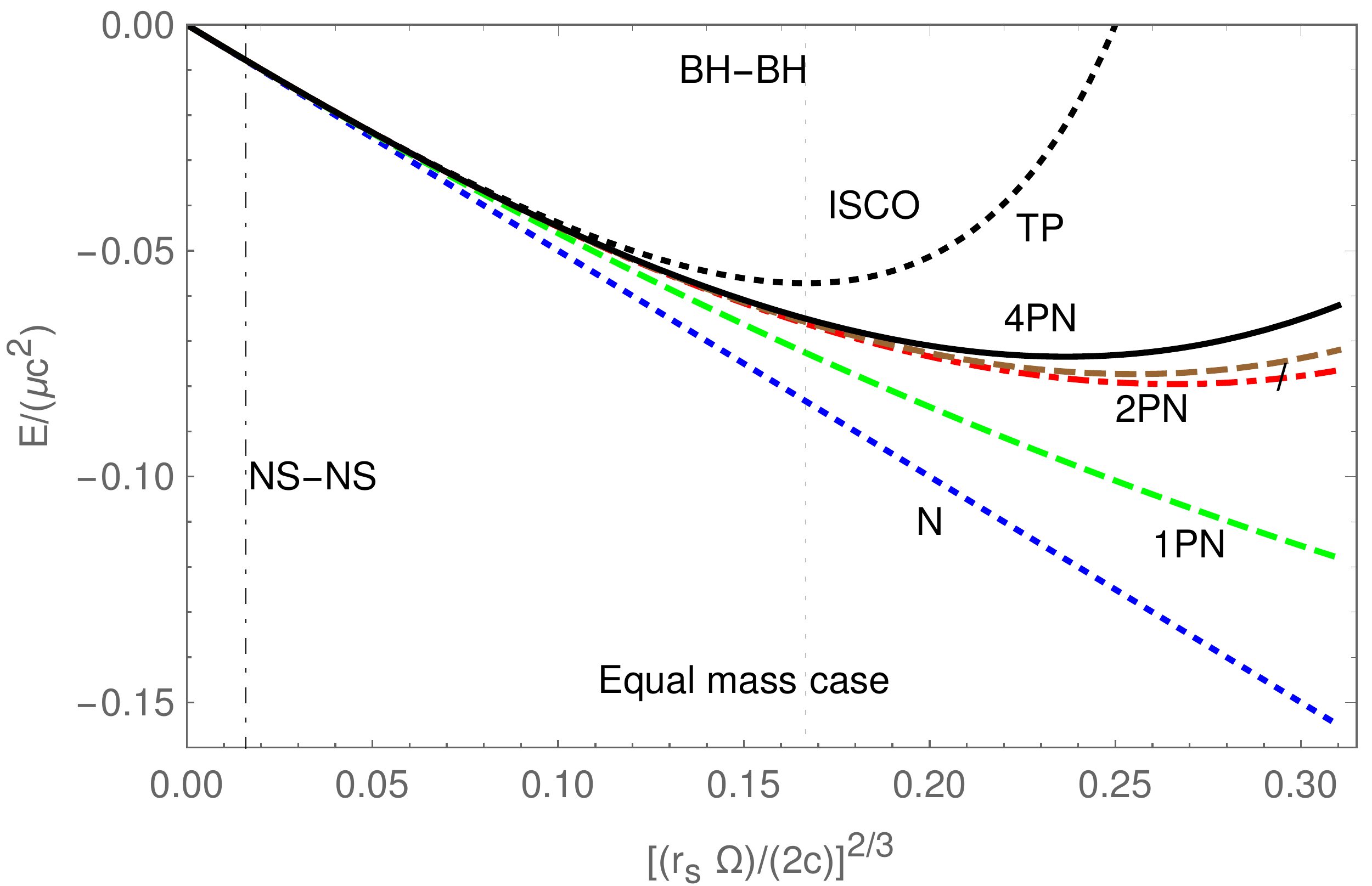}
  \caption[]{\sf The energy of the last stable orbit in the quasi-circular case for equal masses.
  Dotted line: Newtonian case (N); 
  Dashed line: 1PN; 
  Dash-dotted line: 2PN; 
  Upper dashed line: 3PN; 
  Full line: 4PN;
  Upper dotted line: test particle solution (TP). 
  Dashed vertical lines: range for the innermost stable circular orbit (ISCO), \cite{Jaranowski:2013lca}.
  The other vertical lines mark the frequency spectrum for neutron star (NS) and black hole (BH) merging at LIGO.}
  \label{FIG1}
\end{figure}
%--------------------------------------------------------------------------------------------------------
%--------------------------------------------------------------------------------------------------------
\begin{figure}[H]
  \centering
  \hskip-0.8cm
  \includegraphics[width=.8\linewidth]{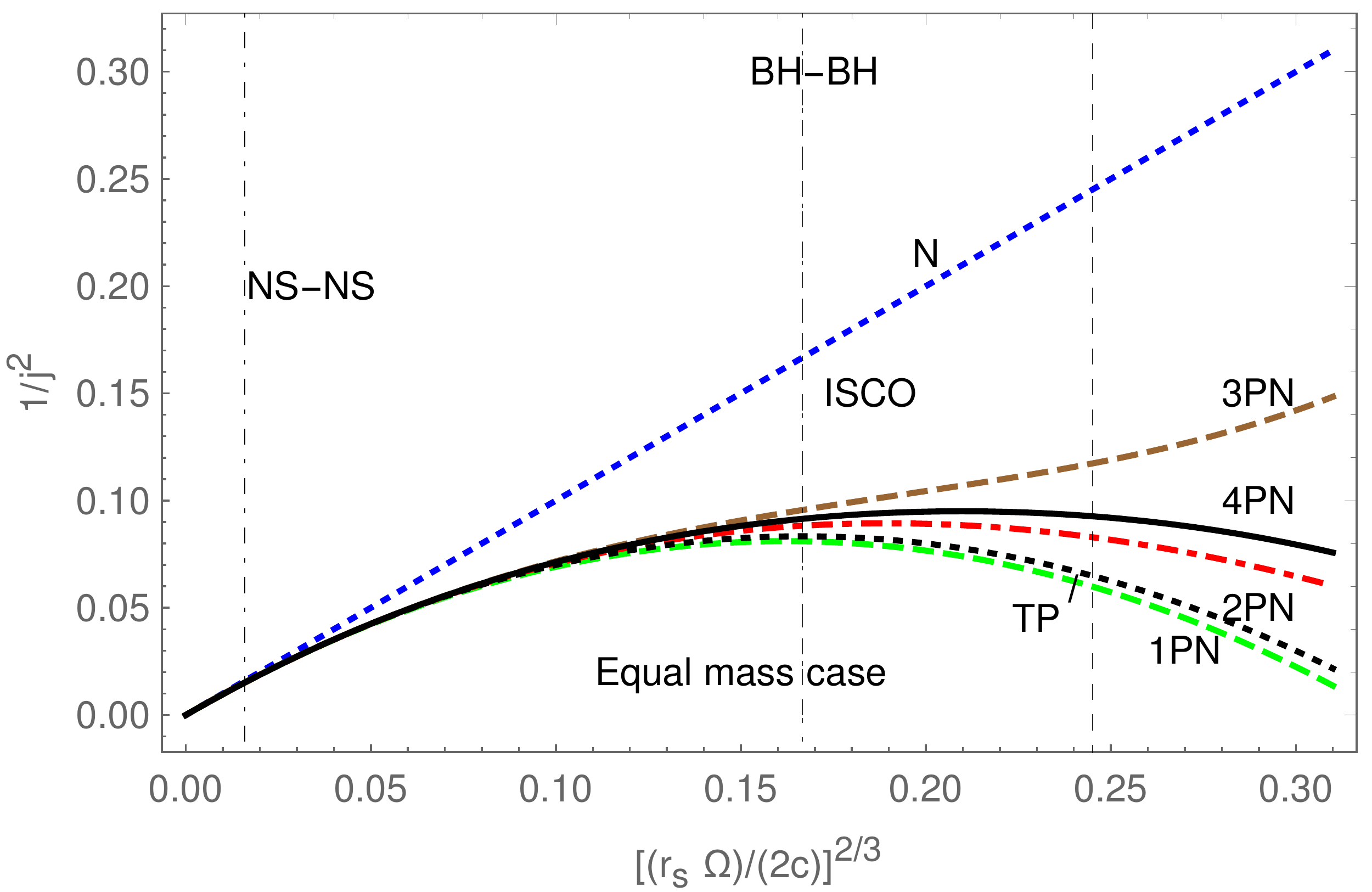}
  \caption[]{\sf The angular momentum $1/j^2$; the labels correspond to Figure~\ref{FIG1}.} 
  \label{FIG2}
\end{figure}
%--------------------------------------------------------------------------------------------------------
The test particle solution reads \cite{Schafer:2018kuf}
%--------------------------------------------------------------------------------------------------------
\begin{eqnarray}
\label{eq:ANGTP}
\frac{1}{j^2(x)} &=& x (1-3x).
\end{eqnarray}
%--------------------------------------------------------------------------------------------------------
Both relations (\ref{eq:ENER},\ref{eq:ANG}) are well--known from the results of other calculations, cf. 
e.g.~\cite{Schafer:2018kuf}. 
The position of the innermost stable circular orbit varies with $\nu$ \cite{Jaranowski:2013lca}. In the limit $\nu 
\rightarrow 
0$ it assumes $x = 1/6$ and in the equal mass case at 4PN it reaches $x \approx 0.245$.

We have obtained agreement with the above relations by explicit calculation, which are illustrated 
in Figures~\ref{FIG1}--\ref{FIG2}. Here $r_s = 2 G_N M/c^2$ denotes the Schwarzschild radius. 
It is evident that even in the region of ISCO at 4PN convergence has
not been reached and even higher order corrections are important. In these Figures we limited ourselves
to the post--Newtonian approach. When extending the $x$-range, further matching to the results of other 
representations becomes necessary, cf. e.g. \cite{MERG}.

The final goal of the post--Newtonian calculations is the prediction of the oscillation signal due to 
gravitational
waves in the detectors. This requires in addition to the Hamiltonian also the luminosity function $L$, cf. e.g. 
\cite{JK}, which currently only allows to use the 3PN approximation on the Hamiltonian side,  which we agree,
cf. e.g.~\cite{JK,Blanchet:2019zlt}.
%--------------------------------------------------------------------------------------------------------
\section{Conclusions}
\label{sec:6}
%--------------------------------------------------------------------------------------------------------

\vspace*{1mm}
\noindent
We have calculated the Hamiltonian of two-body systems in an effective field theory approach to gravity
to the fourth post--Newtonian order in harmonic coordinates using dimensional regularization. Here, starting 
with the third post--Newtonian order divergences in the dimensional regularization parameter $\ep$ occur. 
However, these terms do not affect any observable. These are gauge- and coordinate system artifacts related 
to the regularization process. No renormalization of the observables related to the process under consideration 
is needed. We have also presented pole--free Hamiltonians, obtained after a canonical transformation. Already before 
similar calculations in harmonic coordinates were performed in Refs.~\cite{Bernard:2015njp,Bernard:2017ktp,
Foffa:2019rdf}. The present calculation, however, cannot be compared literally to these, because of probable 
differences in Feynman rules and differences in total time derivatives, which  are partly large contributions. 
We have checked for \cite{Foffa:2019rdf} that we obtain the same equation of motion for the $m$th order Lagrange 
density. Other differences lay in the elimination of the acceleration terms and their time derivatives by forming 
the first order effective Lagrangian. The approach in \cite{Bernard:2017ktp} (and references therein) in harmonic 
coordinates is not an effective field theory approach, but is based on the Fokker action \cite{FOKKER}. However, 
we have shown by an explicit calculation, that the Lagrangian given \cite{Bernard:2017ktp} yield the same result 
for $E(\Omega)$ which we obtain.

We compared to other results obtained working in harmonic coordinates \cite{Bernard:2017ktp,Foffa:2019rdf}, ADM 
coordinates \cite{Damour:2014jta} and EOB coordinates \cite{Bini:2019nra} and demonstrated equivalence on the 
Hamiltonian level by the construction of explicit canonical transformations. We have illustrated our results in 
the case of circular motion for $E(\Omega)$ and  $j(\Omega)$.
We also present the Hamiltonian resulting from harmonic coordinates up to $O(\ep)$ after eliminating the acceleration by a 
shift in $D$ dimensions, which is useful for further higher order calculations.
%--------------------------------------------------------------------------------------------------------
%--------------------------------------------------------------------------------------------------------
\appendix
\section{The 4PN Hamiltonian up to \boldmath $O(\ep)$}
\label{sec:A}
%--------------------------------------------------------------------------------------------------------

\vspace*{1mm}
\noindent
In the following we present the local Hamiltonian up to 4PN including the $O(\ep)$ terms, a central 
result of the present calculation. In its details it also serves as an important input to higher order post--Newtonian
calculations. We use the following abbreviations
%--------------------------------------------------------------------------------------------------------
\begin{eqnarray}
\hat{r} = r \mu_1,~~~~~S_\ep = (4\pi)^\ep \exp[\gamma_E \ep],~~~~~
\tilde{S}_\ep =  \pi^\ep \exp[\gamma_E \ep]
\end{eqnarray}
%--------------------------------------------------------------------------------------------------------
and obtain\footnote{
The polynomials $\left.P_i\right|_{i=1}^{321}$ in the above equations are very lengthy and are given in the 
computer-readable file attached to this paper.}
%--------------------------------------------------------------------------------------------------------
\begin{align}
H_{\text{N}}={}&\frac{p_1^2}{2 m_1}
+\frac{p_2^2}{2 m_2}+\tilde{S}_{\varepsilon} P_{10} \frac{G_N}{r}\,,\\
H_{\text{1PN}}={}&-\frac{p_1^4}{8 m_1^3}
-\frac{p_2^4}{8 m_2^3} + \tilde{S}_{\varepsilon} \bigg\{
        \bigg[
                \varepsilon  \bigg(
                        -\frac{3}{2} n.p_1 n.p_2
                        +\frac{p_1.p_2}{2}
                        +n.p_1 n.p_2 \ln (2)
                        +7 p_1.p_2 \ln (2)
                        \nonumber\\&
                        +n.p_1 n.p_2 \ln(\hat{r})
                        +7 p_1.p_2 \ln(\hat{r})
                        +\frac{P_{168} m_1}{m_2}
                        +\frac{P_{74} m_2}{m_1}
                \bigg)
                +\frac{1}{2} n.p_1 n.p_2
                +\frac{7 p_1.p_2}{2}
                \nonumber\\&
                -\frac{3 m_2 p_1^2}{2 m_1}
                -\frac{3 m_1 p_2^2}{2 m_2}
        \bigg] \frac{G_N}{r}
        +P_{26} \left(\frac{G_N}{r}\right)^2 S_{\varepsilon }
\bigg\}
\,,\\
H_{\text{2PN}}={}&\frac{p_1^6}{16 m_1^5}
+\frac{p_2^6}{16 m_2^5} + \tilde{S}_{\varepsilon} \bigg\{
        \bigg[
                \varepsilon  \bigg(
                        \frac{P_{86}}{m_1^2}
                        +\frac{P_{253} m_1}{m_2^3}
                        +\frac{P_{189}}{m_2^2}
                        +\frac{P_{212}}{m_1 m_2}
                        +\frac{P_{107} m_2}{m_1^3}
                \bigg)
                +\frac{P_{63}}{m_1^2}
                +\frac{P_{153}}{m_2^2}\nonumber\\&
                +\frac{P_{203}}{m_1 m_2}
                +\frac{5 m_2 p_1^4}{8 m_1^3}
                +\frac{5 m_1 p_2^4}{8 m_2^3}
        \bigg] \frac{G_N}{r}
        +\bigg[
                \varepsilon  \bigg(
                        P_{162} m_1
                        +\frac{P_{165} m_1^2}{m_2}
                        +P_{68} m_2
                        +\frac{P_{71} m_2^2}{m_1}
                \bigg)
                \nonumber\\&
                +P_{132} m_1
                +\frac{P_{133} m_1^2}{m_2}
                +P_{48} m_2
                +\frac{P_{49}
                 m_2^2}{m_1}
        \bigg] \left(\frac{G_N}{r}\right)^2 S_{\varepsilon }
        +P_{27} \left(\frac{G_N}{r}\right)^3 S_{\varepsilon }^2
\bigg\}
+\tilde{S}_{\varepsilon}^2 \bigg[
        \varepsilon  \bigg(\nonumber\\&
                P_{173} m_1
                +
                \frac{P_{172} m_1^2}{m_2}
                +P_{78} m_2
                +\frac{P_{77} m_2^2}{m_1}
        \bigg)
        +P_{144} m_1
        +P_{57} m_2
        +\frac{9 m_2^2 p_1^2}{2 m_1}
        +\frac{9 m_1^2 p_2^2}{2 m_2}
\bigg] \left(\frac{G_N}{r}\right)^2\nonumber\\&
+\tilde{S}_{\varepsilon}^3 P_{12} \left(\frac{G_N}{r}\right)^3
\,,\\
H_{\text{3PN}}={}&-\frac{5 p_1^8}{128 m_1^7}
-\frac{5 p_2^8}{128 m_2^7}+\tilde{S}_{\varepsilon}^3 \bigg\{
        \bigg[
                \varepsilon  \bigg(
                        P_{156} m_1^2
                        +\frac{P_{158} m_1^3}{m_2}
                        +P_{157} m_1 m_2
                        +P_{65} m_2^2
                        +\frac{P_{66} m_2^3}{m_1}
                \bigg)
                \nonumber\\&
                +P_{127} m_1^2
                +P_{129} m_1 m_2
                +P_{44} m_2^2
                -\frac{27 m_2^3 p_1^2}{2 m_1}
                -\frac{27 m_1^3 p_2^2}{2 m_2}
        \bigg] \left(\frac{G_N}{r}\right)^3
        +P_{37} \left(\frac{G_N}{r}\right)^4 S_{\varepsilon }
\bigg\}
\nonumber\\&
+\tilde{S}_{\varepsilon}^2 \bigg\{
        \bigg[
                \varepsilon  \bigg(
                        \frac{P_{214}}{m_1}
                        +\frac{P_{252} m_1^2}{m_2^3}
                        +\frac{P_{256} m_1}{m_2^2}
                        +\frac{P_{215}}{m_2}
                        +\frac{P_{109} m_2}{m_1^2}
                        +\frac{P_{106} m_2^2}{m_1^3}
                \bigg)
                +\frac{P_{205}}{m_1}
                +\frac{P_{241} m_1}{m_2^2}
                \nonumber\\&
                +\frac{P_{206}}{m_2}
                +\frac{P_{97} m_2}{m_1^2}
                -\frac{3 m_2^2 p_1^4}{4 m_1^3}
                -\frac{3 m_1^2 p_2^4}{4 m_2^3}
        \bigg] \left(\frac{G_N}{r}\right)^2
        +\bigg[
                \varepsilon  \bigg(
                        P_{184}
                         m_1^2
                        +
                        \frac{P_{181} m_1^3}{m_2}
                        +P_{182} m_1 m_2
                        \nonumber\\&
                        +P_{83} m_2^2
                        +\frac{P_{81} m_2^3}{m_1}
                \bigg)
                +P_{148} m_1^2
                +\frac{P_{145} m_1^3}{m_2}
                +P_{147} m_1 m_2
                +P_{60} m_2^2
                +\frac{P_{58} m_2^3}{m_1}
         \bigg] \left(\frac{G_N}{r}\right)^3 S_{\varepsilon }
                \nonumber\\&
        +P_{39} \left(\frac{G_N}{r}\right)^4 S_{\varepsilon }^2
\bigg\}
+\tilde{S}_{\varepsilon} \bigg\{
        \bigg[
                \varepsilon  \bigg(
                        \frac{P_{113}}{m_1^4}
                        +\frac{P_{283} m_1}{m_2^5}
                        +\frac{P_{260}}{m_2^4}
                        +\frac{P_{273}}{m_1 m_2^3}
                        +\frac{P_{217}}{m_1^2 m_2^2}
                        +\frac{P_{228}}{m_1^3 m_2}
                        \nonumber\\&
                        +\frac{P_{120} m_2}{m_1^5}
                \bigg)
                +\frac{P_{100}}{m_1^4}
                +\frac{P_{244}}{m_2^4}
                +\frac{P_{267}}{m_1 m_2^3}
                +\frac{P_{208}}{m_1^2 m_2^2}
                +\frac{P_{222}}{m_1^3 m_2}
                -\frac{7 m_2 p_1^6}{16 m_1^5}
                -\frac{7 m_1 p_2^6}{16 m_2^5}
        \bigg] \frac{G_N}{r}
        \nonumber\\&
        +\bigg[
                \varepsilon  \bigg(
                        \frac{P_{210}}{m_1}
                        +\frac{P_{250} m_1^2}{m_2^3}
                        +\frac{P_{251} m_1}{m_2^2}
                        +\frac{P_{211}}{m_2}
                        +\frac{P_{105} m_2}{m_1^2}
                        +\frac{P_{104} m_2^2}{m_1^3}
                \bigg)
                +\frac{P_{201}}{m_1}
                +\frac{P_{237} m_1^2}{m_2^3}
                \nonumber\\&
                +\frac{P_{238} m_1}{m_2^2}
                +\frac{P_{202}
                }{m_2}
                +
                \frac{P_{95} m_2}{m_1^2}
                +\frac{P_{94} m_2^2}{m_1^3}
        \bigg] \left(\frac{G_N}{r}\right)^2 S_{\varepsilon }
        +\bigg[
                \pi ^2 P_{292}
                +\varepsilon  \bigg(
                        \pi ^2 P_{295}
                        +P_{193} m_1^2
                        \nonumber\\&
                        +\frac{P_{163} m_1^3}{m_2}
                        +P_{192} m_1 m_2
                        +P_{88} m_2^2
                        +\frac{P_{69} m_2^3}{m_1}
                \bigg)
                +P_{166} m_1^2
                +\frac{P_{135} m_1^3}{m_2}
                +P_{161} m_1 m_2
                \nonumber\\&
                +P_{72} m_2^2
                +\frac{P_{50} m_2^3}{m_1}
        \bigg] \left(\frac{G_N}{r}\right)^3 S_{\varepsilon }^2
        +P_{38} \left(\frac{G_N}{r}\right)^4 S_{\varepsilon }^3
\bigg\}
+\tilde{S}_{\varepsilon}^4 P_{30} \left(\frac{G_N}{r}\right)^4\,,\\
H_{\text{4PN}}={}&
\frac{7 p_1^{10}}{256 m_1^9}
+\frac{7 p_2^{10}}{256 m_2^9}
+\tilde{S}_{\varepsilon}^4 \bigg\{
        \bigg[
                \varepsilon  \bigg(
                        P_{187} m_1^3
                        +\frac{P_{185} m_1^4}{m_2}
                        +P_{188} m_1^2 m_2
                        +P_{186} m_1 m_2^2
                        +P_{85} m_2^3
                        \nonumber\\&
                        +\frac{P_{84} m_2^4}{m_1}
                \bigg)
                +P_{150} m_1^3
                +P_{151} m_1^2 m_2
                +P_{149} m_1 m_2^2
                +P_{61} m_2^3
                +\frac{81 m_2^4 p_1^2}{2 m_1}
                +\frac{81 m_1^4 p_2^2}{2 m_2}
        \bigg]
        \nonumber\\&
        \left(\frac{G_N}{r}\right)^4
        +P_{40} \left(\frac{G_N}{r}\right)^5 S_{\varepsilon }
\bigg\}
+\tilde{S}_{\varepsilon}^3 \bigg\{
        \bigg[
                \varepsilon  \bigg(
                        \frac{225}{4} (n.p_1)^3 n.p_2
                        -\frac{92297}{30}
                         (n.p_1)^2 (n.p_2)^2
                         \nonumber\\&
                         +\frac{225}{4} n.p_1 (n.p_2)^3
                        +\frac{53}{2} (n.p_1)^2 p_1.p_2
                        +\frac{2945849 n.p_1 n.p_2 p_1.p_2}{1200}
                        +\frac{53}{2} (n.p_2)^2 p_1.p_2
                        \nonumber\\&
                        -\frac{496543 (p_1.p_2)^2}{1800}
                        -\frac{177}{2} (n.p_1)^3 n.p_2 \ln (2)
                        +2473 (n.p_1)^2 (n.p_2)^2 \ln (2)
                        \nonumber\\&
                        -\frac{177}{2} n.p_1 (n.p_2)^3 \ln (2)
                        -\frac{1101}{4} (n.p_1)^2 p_1.p_2 \ln (2)
                        -\frac{103357}{40} n.p_1 n.p_2 p_1.p_2 \ln (2)
                        \nonumber\\&
                        -\frac{1101}{4} (n.p_2)^2 p_1.p_2 \ln (2)
                        -\frac{75359}{40} (p_1.p_2)^2 \ln (2)
                        -\frac{177}{2} (n.p_1)^3 n.p_2 \ln(\hat{r})
                        \nonumber\\&
                        +2473 (n.p_1)^2 (n.p_2)^2 \ln(\hat{r})
                        -\frac{177}{2} n.p_1 (n.p_2)^3 \ln(\hat{r})
                        -\frac{1101}{4} (n.p_1)^2 p_1.p_2 \ln(\hat{r})
                        \nonumber\\&
                        -\frac{103357}{40} n.p_1 n.p_2 p_1.p_2 \ln(\hat{r})
                        -\frac{1101}{4} (n.p_2)^2 p_1.p_2 \ln(\hat{r})
                        -\frac{75359}{40} (p_1.p_2)^2 \ln(\hat{r})
                        \nonumber\\&
                        +\frac{P_{247} m_1^3}{m_2^3}
                        +\frac{P_{246}
                         m_1^2}{m_2^2}
                        +
                        \frac{P_{245} m_1}{m_2}
                        +\frac{P_{209} m_2}{m_1}
                        +\frac{P_{101} m_2^2}{m_1^2}
                        +\frac{P_{102} m_2^3}{m_1^3}
                        -\frac{17525}{144} n.p_1 n.p_2 p_1^2
                        \nonumber\\&
                        +\frac{1282037 (n.p_2)^2 p_1^2}{2400}
                        +\frac{25133}{432} p_1.p_2 p_1^2
                        +\frac{289}{2} n.p_1 n.p_2 \ln (2) p_1^2
                        -\frac{24221}{80} (n.p_2)^2 \ln (2) p_1^2
                        \nonumber\\&
                        +\frac{3490}{3} p_1.p_2 \ln (2) p_1^2
                        +\frac{289}{2} n.p_1 n.p_2 \ln(\hat{r}) p_1^2
                        -\frac{24221}{80} (n.p_2)^2 \ln(\hat{r}) p_1^2
                        \nonumber\\&
                        +\frac{3490}{3} p_1.p_2 \ln(\hat{r}) p_1^2
                        +\frac{1282037 (n.p_1)^2 p_2^2}{2400}
                        -\frac{17525}{144} n.p_1 n.p_2 p_2^2
                        +\frac{25133}{432} p_1.p_2 p_2^2
                        \nonumber\\&
                        -\frac{24221}{80} (n.p_1)^2 \ln (2) p_2^2
                        +\frac{289}{2} n.p_1 n.p_2 \ln (2) p_2^2
                        +\frac{3490}{3} p_1.p_2 \ln (2) p_2^2
                        \nonumber\\&
                        -\frac{24221}{80} (n.p_1)^2 \ln(\hat{r}) p_2^2
                        +\frac{289}{2} n.p_1 n.p_2 \ln(\hat{r}) p_2^2
                        +\frac{3490}{3} p_1.p_2 \ln(\hat{r}) p_2^2
                        -\frac{145817}{900} p_1^2 p_2^2
                        \nonumber\\&
                        -\frac{49597}{40} \ln (2) p_1^2 p_2^2
                        -\frac{49597}{40} \ln(\hat{r}) p_1^2 p_2^2
                \bigg)
                -
                \frac{59}{4} (n.p_1)^3 n.p_2
                +\frac{2473}{6} (n.p_1)^2 (n.p_2)^2
                \nonumber\\&
                -\frac{59}{4} n.p_1 (n.p_2)^3
                -\frac{367}{8} (n.p_1)^2 p_1.p_2
                -\frac{103357}{240} n.p_1 n.p_2 p_1.p_2
                -\frac{367}{8} (n.p_2)^2 p_1.p_2
                \nonumber\\&
                -\frac{75359}{240} (p_1.p_2)^2
                +\frac{P_{234} m_1^2}{m_2^2}
                +\frac{P_{233} m_1}{m_2}
                +\frac{P_{200} m_2}{m_1}
                +\frac{P_{91} m_2^2}{m_1^2}
                +\frac{289}{12} n.p_1 n.p_2 p_1^2
                \nonumber\\&
                -\frac{24221}{480} (n.p_2)^2 p_1^2
                +\frac{1745}{9} p_1.p_2 p_1^2
                -\frac{45 m_2^3 p_1^4}{4 m_1^3}
                -\frac{24221}{480} (n.p_1)^2 p_2^2
                +\frac{289}{12} n.p_1 n.p_2 p_2^2
                \nonumber\\&
                +\frac{1745}{9} p_1.p_2 p_2^2
                -\frac{49597}{240} p_1^2 p_2^2
                -\frac{45 m_1^3 p_2^4}{4 m_2^3}
        \bigg] \left(\frac{G_N}{r}\right)^3
        +\bigg[
                \varepsilon  \bigg(
                        \pi ^2 P_{291}
                        +P_{190} m_1^3
                        +\frac{P_{155} m_1^4}{m_2}
                        \nonumber\\&
                        +P_{191} m_1^2 m_2
                        +P_{154} m_1 m_2^2
                        +P_{87} m_2^3
                        +\frac{P_{64} m_2^4}{m_1}
                \bigg)
                +P_{159} m_1^3
                +\frac{P_{128} m_1^4}{m_2}
                +P_{160} m_1^2 m_2
                \nonumber\\&
                +P_{126}
                 m_1 m_2^2
                +P_{67} m_2^3
                +
                \frac{P_{45} m_2^4}{m_1}
        \bigg] \left(\frac{G_N}{r}\right)^4 S_{\varepsilon }
        +P_{43} \left(\frac{G_N}{r}\right)^5 S_{\varepsilon }^2
\bigg\}
+\tilde{S}_{\varepsilon}^2 \bigg\{
        \bigg[
                \nonumber\\&
                \varepsilon  \bigg(
                        \frac{P_{226}}{m_1^3}
                        +\frac{P_{285} m_1^2}{m_2^5}
                        +\frac{P_{286} m_1}{m_2^4}
                        +\frac{P_{271}}{m_2^3}
                        +\frac{P_{274}}{m_1 m_2^2}
                        +\frac{P_{229}}{m_1^2 m_2}
                        +\frac{P_{123} m_2}{m_1^4}
                        +\frac{P_{122} m_2^2}{m_1^5}
                \bigg)
                +\frac{P_{220}}{m_1^3}
                \nonumber\\&
                +\frac{P_{280} m_1}{m_2^4}
                +\frac{P_{265}}{m_2^3}
                +\frac{P_{268}}{m_1 m_2^2}
                +\frac{P_{223}}{m_1^2 m_2}
                +\frac{P_{117} m_2}{m_1^4}
                +\frac{23 m_2^2 p_1^6}{16 m_1^5}
                +\frac{23 m_1^2 p_2^6}{16 m_2^5}
        \bigg] \left(\frac{G_N}{r}\right)^2
        +\bigg[
                \varepsilon  \bigg(
                        \nonumber\\&
                        \pi ^2 \bigg(
                                \frac{P_{236} m_1}{m_2}
                                +\frac{P_{93} m_2}{m_1}
                                -\frac{33}{8} n.p_1 n.p_2 p_1^2
                                +\frac{11}{8} p_1.p_2 p_1^2
                                -\frac{33}{8} n.p_1 n.p_2 p_2^2
                                +\frac{11}{8} p_1.p_2 p_2^2
                        \bigg)
                        \nonumber\\&
                        -\frac{7219}{18} (n.p_1)^3 n.p_2
                        +\frac{391051}{90} (n.p_1)^2 (n.p_2)^2
                        -\frac{7219}{18} n.p_1 (n.p_2)^3
                        +\frac{568}{3} (n.p_1)^2 p_1.p_2
                        \nonumber\\&
                        -\frac{188107}{90} n.p_1 n.p_2 p_1.p_2
                        +
                        \frac{568}{3} (n.p_2)^2 p_1.p_2
                        +\frac{22691}{150} (p_1.p_2)^2
                        +278 (n.p_1)^3 n.p_2 \ln (2)
                        \nonumber\\&
                        -\frac{8144}{3} (n.p_1)^2 (n.p_2)^2 \ln (2)
                        +278 n.p_1 (n.p_2)^3 \ln (2)
                        -24 (n.p_1)^2 p_1.p_2 \ln (2)
                        \nonumber\\&
                        +\frac{3830}{3} n.p_1 n.p_2 p_1.p_2 \ln (2)
                        -24 (n.p_2)^2 p_1.p_2 \ln (2)
                        +\frac{6518}{5} (p_1.p_2)^2 \ln (2)
                        \nonumber\\&
                        +417 (n.p_1)^3 n.p_2 \ln(\hat{r})
                        -4072 (n.p_1)^2 (n.p_2)^2 \ln(\hat{r})
                        +417 n.p_1 (n.p_2)^3 \ln(\hat{r})
                        \nonumber\\&
                        -36 (n.p_1)^2 p_1.p_2 \ln(\hat{r})
                        +1915 n.p_1 n.p_2 p_1.p_2 \ln(\hat{r})
                        -36 (n.p_2)^2 p_1.p_2 \ln(\hat{r})
                        \nonumber\\&
                        +\frac{9777}{5} (p_1.p_2)^2 \ln(\hat{r})
                        +\frac{P_{258} m_1^3}{m_2^3}
                        +\frac{P_{259} m_1^2}{m_2^2}
                        +\frac{P_{261} m_1}{m_2}
                        +\frac{P_{216} m_2}{m_1}
                        +\frac{P_{112} m_2^2}{m_1^2}
                        +\frac{P_{111} m_2^3}{m_1^3}
                        \nonumber\\&
                        -\frac{9787}{144} n.p_1 n.p_2 p_1^2
                        -\frac{113167}{360} (n.p_2)^2 p_1^2
                        -\frac{17963}{432} p_1.p_2 p_1^2
                        +\frac{1475}{12} n.p_1 n.p_2 \ln (2) p_1^2
                        \nonumber\\&
                        -\frac{29}{6} (n.p_2)^2 \ln (2) p_1^2
                        -\frac{26015}{36} p_1.p_2 \ln (2) p_1^2
                        -44 n.p_1 n.p_2 \ln ^2(2) p_1^2
                        +\frac{44}{3} p_1.p_2 \ln ^2(2) p_1^2
                        \nonumber\\&
                        +\frac{1475}{8} n.p_1 n.p_2 \ln(\hat{r}) p_1^2
                        -\frac{29}{4} (n.p_2)^2 \ln(\hat{r}) p_1^2
                        -\frac{26015}{24} p_1.p_2 \ln(\hat{r}) p_1^2
                        \nonumber\\&
                        -132 n.p_1 n.p_2 \ln (2) \ln(\hat{r}) p_1^2
                        +44 p_1.p_2 \ln (2) \ln(\hat{r}) p_1^2
                        -99 n.p_1 n.p_2 \ln ^2(\hat{r}) p_1^2
                        \nonumber\\&
                        +33 p_1.p_2 \ln ^2(\hat{r}) p_1^2
                        -\frac{113167}{360} (n.p_1)^2 p_2^2
                        -\frac{9787}{144} n.p_1 n.p_2 p_2^2
                        -\frac{17963}{432} p_1.p_2 p_2^2
                        \nonumber\\&
                        -\frac{29}{6} (n.p_1)^2 \ln (2) p_2^2
                        +\frac{1475}{12} n.p_1 n.p_2 \ln (2) p_2^2
                        -\frac{26015}{36} p_1.p_2 \ln (2) p_2^2
                        \nonumber\\&
                        -44 n.p_1 n.p_2 \ln ^2(2) p_2^2
                        +\frac{44}{3} p_1.p_2 \ln ^2(2) p_2^2
                        -\frac{29}{4} (n.p_1)^2 \ln(\hat{r}) p_2^2
                        \nonumber\\&
                        +\frac{1475}{8} n.p_1 n.p_2 \ln(\hat{r}) p_2^2
                        -\frac{26015}{24} p_1.p_2 \ln(\hat{r}) p_2^2
                        -132 n.p_1 n.p_2 \ln (2) \ln(\hat{r}) p_2^2
                        \nonumber\\&
                        +44 p_1.p_2 \ln (2) \ln(\hat{r}) p_2^2
                        -99 n.p_1 n.p_2 \ln ^2(\hat{r}) p_2^2
                        +33 p_1.p_2 \ln ^2(\hat{r}) p_2^2
                        +\frac{7237}{225} p_1^2 p_2^2
                        \nonumber\\&
                        +\frac{13232}{15} \ln (2) p_1^2 p_2^2
                        +\frac{6616}{5} \ln(\hat{r}) p_1^2 p_2^2
                \bigg)
                +\frac{139}{2} (n.p_1)^3 n.p_2
                -\frac{2036}{3} (n.p_1)^2 (n.p_2)^2
                \nonumber\\&
                +\frac{139}{2} n.p_1 (n.p_2)^3
                -6 (n.p_1)^2 p_1.p_2
                +\frac{1915}{6} n.p_1 n.p_2 p_1.p_2
                -6 (n.p_2)^2 p_1.p_2
                +\frac{3259}{10} (p_1.p_2)^2
                \nonumber\\&
                +\frac{P_{242} m_1^3}{m_2^3}
                +\frac{P_{243} m_1^2}{m_2^2}
                +\frac{P_{249} m_1}{m_2}
                +\frac{P_{207} m_2}{m_1}
                +\frac{P_{99} m_2^2}{m_1^2}
                +\frac{P_{98} m_2^3}{m_1^3}
                +\frac{1475}{48} n.p_1 n.p_2 p_1^2
                \nonumber\\&
                -\frac{29}{24} (n.p_2)^2 p_1^2
                -\frac{26015}{144} p_1.p_2 p_1^2
                -22 n.p_1 n.p_2 \ln (2) p_1^2
                +\frac{22}{3} p_1.p_2 \ln (2) p_1^2
                \nonumber\\&
                -33 n.p_1 n.p_2 \ln(\hat{r}) p_1^2
                +11 p_1.p_2 \ln(\hat{r}) p_1^2
                -\frac{29}{24} (n.p_1)^2 p_2^2
                +\frac{1475}{48} n.p_1 n.p_2 p_2^2
                \nonumber\\&
                -\frac{26015}{144} p_1.p_2 p_2^2
                -22 n.p_1 n.p_2 \ln (2) p_2^2
                +\frac{22}{3} p_1.p_2 \ln (2) p_2^2
                -33 n.p_1 n.p_2 \ln(\hat{r}) p_2^2
                \nonumber\\&
                +11 p_1.p_2 \ln(\hat{r}) p_2^2
                +\frac{3308}{15} p_1^2 p_2^2
        \bigg] \left(\frac{G_N}{r}\right)^3 S_{\varepsilon }
        +\bigg[
                \pi ^2 P_{294}
                +\varepsilon  \bigg(
                        \pi ^2 P_{297}
                        +P_{199} m_1^3
                        +
                        \frac{P_{183} m_1^4}{m_2}
                        \nonumber\\&
                        +P_{198} m_1^2 m_2
                        +P_{197} m_1 m_2^2
                        +P_{90} m_2^3
                        +\frac{P_{82} m_2^4}{m_1}
                \bigg)
                +P_{179} m_1^3
                +\frac{P_{146} m_1^4}{m_2}
                +P_{178} m_1^2 m_2
                \nonumber\\&
                +P_{177} m_1 m_2^2
                +P_{80} m_2^3
                +\frac{P_{59} m_2^4}{m_1}
        \bigg] \left(\frac{G_N}{r}\right)^4 S_{\varepsilon }^2
        +P_{42} \left(\frac{G_N}{r}\right)^5 S_{\varepsilon }^3
\bigg\}
+\tilde{S}_{\varepsilon} \bigg\{
        \bigg[
                \nonumber\\&
                \varepsilon  \bigg(
                        \frac{P_{124}}{m_1^6}
                        +\frac{P_{290} m_1}{m_2^7}
                        +\frac{P_{287}}{m_2^6}
                        +\frac{P_{289}}{m_1 m_2^5}
                        +\frac{P_{275}}{m_1^2 m_2^4}
                        +\frac{P_{277}}{m_1^3 m_2^3}
                        +\frac{P_{230}}{m_1^4 m_2^2}
                        +\frac{P_{232}}{m_1^5 m_2}
                        +\frac{P_{125} m_2}{m_1^7}
                \bigg)
                \nonumber\\&
                +\frac{P_{118}}{m_1^6}
                +\frac{P_{281}}{m_2^6}
                +\frac{P_{288}}{m_1 m_2^5}
                +\frac{P_{269}}{m_1^2 m_2^4}
                +\frac{P_{276}}{m_1^3 m_2^3}
                +\frac{P_{224}}{m_1^4 m_2^2}
                +\frac{P_{231}}{m_1^5 m_2}
                +\frac{45 m_2 p_1^8}{128 m_1^7}
                +\frac{45 m_1 p_2^8}{128 m_2^7}
        \bigg] \frac{G_N}{r}
        \nonumber\\&
        +\bigg[
                \varepsilon  \bigg(
                        \frac{P_{227}}{m_1^3}
                        +\frac{P_{284} m_1^2}{m_2^5}
                        +\frac{P_{282} m_1}{m_2^4}
                        +\frac{P_{272}
                        }{m_2^3}
                        +
                        \frac{P_{270}}{m_1 m_2^2}
                        +\frac{P_{225}}{m_1^2 m_2}
                        +\frac{P_{119} m_2}{m_1^4}
                        +\frac{P_{121} m_2^2}{m_1^5}
                \bigg)
                \nonumber\\&
                +\frac{P_{221}}{m_1^3}
                +\frac{P_{279} m_1^2}{m_2^5}
                +\frac{P_{278} m_1}{m_2^4}
                +\frac{P_{266}}{m_2^3}
                +\frac{P_{264}}{m_1 m_2^2}
                +\frac{P_{219}}{m_1^2 m_2}
                +\frac{P_{115} m_2}{m_1^4}
                +\frac{P_{116} m_2^2}{m_1^5}
        \bigg] \left(\frac{G_N}{r}\right)^2 S_{\varepsilon }
        \nonumber\\&
        +\bigg[
                \pi ^2 \bigg(
                        \frac{15945}{512} (n.p_1)^2 (n.p_2)^2
                        -\frac{1635}{128} n.p_1 n.p_2 p_1.p_2
                        +\frac{27}{256} (p_1.p_2)^2
                        +\frac{P_{239} m_1^2}{m_2^2}
                        +\frac{P_{152} m_1}{m_2}
                        \nonumber\\&
                        +\frac{P_{62} m_2}{m_1}
                        +\frac{P_{96} m_2^2}{m_1^2}
                        -\frac{1215}{512} (n.p_2)^2 p_1^2
                        -\frac{1215}{512} (n.p_1)^2 p_2^2
                        -\frac{253}{512} p_1^2 p_2^2
                \bigg)
                +\varepsilon  \bigg(
                        \pi ^2 \bigg(
                                \nonumber\\&
                                -35 (n.p_1)^3 n.p_2
                                -\frac{174061}{512} (n.p_1)^2 (n.p_2)^2
                                -35 n.p_1 (n.p_2)^3
                                +\frac{97}{5} (n.p_1)^2 p_1.p_2
                                \nonumber\\&
                                +\frac{19315}{128} n.p_1 n.p_2 p_1.p_2
                                +\frac{97}{5} (n.p_2)^2 p_1.p_2
                                -
                                \frac{21471 (p_1.p_2)^2}{1280}
                                +\frac{47835}{256} (n.p_1)^2 (n.p_2)^2 \ln (2)
                                \nonumber\\&
                                -\frac{4905}{64} n.p_1 n.p_2 p_1.p_2 \ln (2)
                                +\frac{81}{128} (p_1.p_2)^2 \ln (2)
                                +\frac{47835}{256} (n.p_1)^2 (n.p_2)^2 \ln(\hat{r})
                                \nonumber\\&
                                -\frac{4905}{64} n.p_1 n.p_2 p_1.p_2 \ln(\hat{r})
                                +\frac{81}{128} (p_1.p_2)^2 \ln(\hat{r})
                                +\frac{P_{254} m_1^2}{m_2^2}
                                +\frac{P_{240} m_1}{m_2}
                                +\frac{P_{204} m_2}{m_1}
                                \nonumber\\&
                                +\frac{P_{108} m_2^2}{m_1^2}
                                +\frac{319}{20} n.p_1 n.p_2 p_1^2
                                +\frac{43021 (n.p_2)^2 p_1^2}{2560}
                                -\frac{287}{60} p_1.p_2 p_1^2
                                -\frac{3645}{256} (n.p_2)^2 \ln (2) p_1^2
                                \nonumber\\&
                                -\frac{3645}{256} (n.p_2)^2 \ln(\hat{r}) p_1^2
                                +\frac{43021 (n.p_1)^2 p_2^2}{2560}
                                +\frac{319}{20} n.p_1 n.p_2 p_2^2
                                -\frac{287}{60} p_1.p_2 p_2^2
                                \nonumber\\&
                                -\frac{3645}{256} (n.p_1)^2 \ln (2) p_2^2
                                -\frac{3645}{256} (n.p_1)^2 \ln(\hat{r}) p_2^2
                                +\frac{8827 p_1^2 p_2^2}{1536}
                                -\frac{759}{256} \ln (2) p_1^2 p_2^2
                                \nonumber\\&
                                -\frac{759}{256} \ln(\hat{r}) p_1^2 p_2^2
                        \bigg)
                        -\frac{125231}{90} (n.p_1)^3 n.p_2
                        +\frac{160115}{24} (n.p_1)^2 (n.p_2)^2
                        -\frac{125231}{90} n.p_1 (n.p_2)^3
                        \nonumber\\&
                        +\frac{1818287 (n.p_1)^2 p_1.p_2}{3375}
                        -\frac{263129}{45} n.p_1 n.p_2 p_1.p_2
                        +\frac{1818287 (n.p_2)^2 p_1.p_2}{3375}
                        \nonumber\\&
                        +\frac{12737819 (p_1.p_2)^2}{13500}
                        +\frac{2215}{3} (n.p_1)^3 n.p_2 \ln (2)
                        -\frac{168443}{60} (n.p_1)^2 (n.p_2)^2 \ln (2)
                        \nonumber\\&
                        +\frac{2215}{3} n.p_1 (n.p_2)^3 \ln (2)
                        -\frac{75172}{225} (n.p_1)^2 p_1.p_2 \ln (2)
                        +\frac{38974}{15} n.p_1 n.p_2 p_1.p_2 \ln (2)
                        \nonumber\\&
                        -\frac{75172}{225} (n.p_2)^2 p_1.p_2 \ln (2)
                        -\frac{282017}{450} (p_1.p_2)^2 \ln (2)
                        -\frac{280}{3} (n.p_1)^3 n.p_2 \ln ^2(2)
                        \nonumber\\&
                        +300 (n.p_1)^2 (n.p_2)^2 \ln ^2(2)
                        -\frac{280}{3} n.p_1 (n.p_2)^3 \ln ^2(2)
                        +\frac{776}{15} (n.p_1)^2 p_1.p_2 \ln ^2(2)
                        \nonumber\\&
                        -176 n.p_1 n.p_2 p_1.p_2 \ln ^2(2)
                        +\frac{776}{15} (n.p_2)^2 p_1.p_2 \ln ^2(2)
                        +\frac{56}{3} (p_1.p_2)^2 \ln ^2(2)
                        \nonumber\\&
                        +2215 (n.p_1)^3 n.p_2 \ln(\hat{r})
                        -\frac{168443}{20} (n.p_1)^2 (n.p_2)^2 \ln(\hat{r})
                        +2215 n.p_1 (n.p_2)^3 \ln(\hat{r})
                        \nonumber\\&
                        -\frac{75172}{75} (n.p_1)^2 p_1.p_2 \ln(\hat{r})
                        +\frac{38974}{5} n.p_1 n.p_2 p_1.p_2 \ln(\hat{r})
                        -\frac{75172}{75} (n.p_2)^2 p_1.p_2 \ln(\hat{r})
                        \nonumber\\&
                        -\frac{282017}{150} (p_1.p_2)^2 \ln(\hat{r})
                        -560 (n.p_1)^3 n.p_2 \ln (2) \ln(\hat{r})
                        \nonumber\\&
                        +1800 (n.p_1)^2 (n.p_2)^2 \ln (2) \ln(\hat{r})
                        -560 n.p_1 (n.p_2)^3 \ln (2) \ln(\hat{r})
                        \nonumber\\&
                        +\frac{1552}{5} (n.p_1)^2 p_1.p_2 \ln (2) \ln(\hat{r})
                        -1056 n.p_1 n.p_2 p_1.p_2 \ln (2) \ln(\hat{r})
                        \nonumber\\&
                        +\frac{1552}{5} (n.p_2)^2 p_1.p_2 \ln (2) \ln(\hat{r})
                        +112 (p_1.p_2)^2 \ln (2) \ln(\hat{r})
                        -840 (n.p_1)^3 n.p_2 \ln ^2(\hat{r})
                        \nonumber\\&
                        +2700 (n.p_1)^2 (n.p_2)^2 \ln ^2(\hat{r})
                        -840 n.p_1 (n.p_2)^3 \ln ^2(\hat{r})
                        +\frac{2328}{5} (n.p_1)^2 p_1.p_2 \ln ^2(\hat{r})
                        \nonumber\\&
                        -1584 n.p_1 n.p_2 p_1.p_2 \ln ^2(\hat{r})
                        +\frac{2328}{5} (n.p_2)^2 p_1.p_2 \ln ^2(\hat{r})
                        +168 (p_1.p_2)^2 \ln ^2(\hat{r})
                        \nonumber\\&
                        +\frac{P_{248} m_1^3}{m_2^3}
                        +\frac{P_{263} m_1^2}{m_2^2}
                        +\frac{P_{262} m_1}{m_2}
                        +\frac{P_{218} m_2}{m_1}
                        +\frac{P_{114} m_2^2}{m_1^2}
                        +\frac{P_{103} m_2^3}{m_1^3}
                        +\frac{2713489 n.p_1 n.p_2 p_1^2}{5400}
                        \nonumber\\&
                        -\frac{13943393 (n.p_2)^2 p_1^2}{9000}
                        -\frac{605699 p_1.p_2 p_1^2}{9000}
                        -\frac{284573}{900} n.p_1 n.p_2 \ln (2) p_1^2
                        \nonumber\\&
                        +\frac{220249}{300} (n.p_2)^2 \ln (2) p_1^2
                        +\frac{97787}{900} p_1.p_2 \ln (2) p_1^2
                        +\frac{638}{15} n.p_1 n.p_2 \ln ^2(2) p_1^2
                        \nonumber\\&
                        -92 (n.p_2)^2 \ln ^2(2) p_1^2
                        -\frac{574}{45} p_1.p_2 \ln ^2(2) p_1^2
                        -\frac{284573}{300} n.p_1 n.p_2 \ln(\hat{r}) p_1^2
                        \nonumber\\&
                        +\frac{220249}{100} (n.p_2)^2 \ln(\hat{r}) p_1^2
                        +\frac{97787}{300} p_1.p_2 \ln(\hat{r}) p_1^2
                        +\frac{1276}{5} n.p_1 n.p_2 \ln (2) \ln(\hat{r}) p_1^2
                        \nonumber\\&
                        -552 (n.p_2)^2 \ln (2) \ln(\hat{r}) p_1^2
                        -\frac{1148}{15} p_1.p_2 \ln (2) \ln(\hat{r}) p_1^2
                        +\frac{1914}{5} n.p_1 n.p_2 \ln ^2(\hat{r}) p_1^2
                        \nonumber\\&
                        -828 (n.p_2)^2 \ln ^2(\hat{r}) p_1^2
                        -\frac{574}{5} p_1.p_2 \ln ^2(\hat{r}) p_1^2
                        -
                        \frac{13943393 (n.p_1)^2 p_2^2}{9000}
                        \nonumber\\&
                        +\frac{2713489 n.p_1 n.p_2 p_2^2}{5400}
                        -\frac{605699 p_1.p_2 p_2^2}{9000}
                        +\frac{220249}{300} (n.p_1)^2 \ln (2) p_2^2
                        \nonumber\\&
                        -\frac{284573}{900} n.p_1 n.p_2 \ln (2) p_2^2
                        +\frac{97787}{900} p_1.p_2 \ln (2) p_2^2
                        -92 (n.p_1)^2 \ln ^2(2) p_2^2
                        \nonumber\\&
                        +\frac{638}{15} n.p_1 n.p_2 \ln ^2(2) p_2^2
                        -\frac{574}{45} p_1.p_2 \ln ^2(2) p_2^2
                        +\frac{220249}{100} (n.p_1)^2 \ln(\hat{r}) p_2^2
                        \nonumber\\&
                        -\frac{284573}{300} n.p_1 n.p_2 \ln(\hat{r}) p_2^2
                        +\frac{97787}{300} p_1.p_2 \ln(\hat{r}) p_2^2
                        -552 (n.p_1)^2 \ln (2) \ln(\hat{r}) p_2^2
                        \nonumber\\&
                        +\frac{1276}{5} n.p_1 n.p_2 \ln (2) \ln(\hat{r}) p_2^2
                        -\frac{1148}{15} p_1.p_2 \ln (2) \ln(\hat{r}) p_2^2
                        -828 (n.p_1)^2 \ln ^2(\hat{r}) p_2^2
                        \nonumber\\&
                        +\frac{1914}{5} n.p_1 n.p_2 \ln ^2(\hat{r}) p_2^2
                        -\frac{574}{5} p_1.p_2 \ln ^2(\hat{r}) p_2^2
                        +\frac{2824193 p_1^2 p_2^2}{5400}
                        -\frac{3685}{12} \ln (2) p_1^2 p_2^2
                        \nonumber\\&
                        +\frac{124}{3} \ln ^2(2) p_1^2 p_2^2
                        -\frac{3685}{4} \ln(\hat{r}) p_1^2 p_2^2
                        +248 \ln (2) \ln(\hat{r}) p_1^2 p_2^2
                        +372 \ln ^2(\hat{r}) p_1^2 p_2^2
                \bigg)
                \nonumber\\&
                +
                \frac{2215}{6} (n.p_1)^3 n.p_2
                -\frac{168443}{120} (n.p_1)^2 (n.p_2)^2
                +\frac{2215}{6} n.p_1 (n.p_2)^3
                -\frac{37586}{225} (n.p_1)^2 p_1.p_2
                \nonumber\\&
                +\frac{19487}{15} n.p_1 n.p_2 p_1.p_2
                -\frac{37586}{225} (n.p_2)^2 p_1.p_2
                -\frac{282017}{900} (p_1.p_2)^2
                -\frac{280}{3} (n.p_1)^3 n.p_2 \ln (2)
                \nonumber\\&
                +300 (n.p_1)^2 (n.p_2)^2 \ln (2)
                -\frac{280}{3} n.p_1 (n.p_2)^3 \ln (2)
                +\frac{776}{15} (n.p_1)^2 p_1.p_2 \ln (2)
                \nonumber\\&
                -176 n.p_1 n.p_2 p_1.p_2 \ln (2)
                +\frac{776}{15} (n.p_2)^2 p_1.p_2 \ln (2)
                +\frac{56}{3} (p_1.p_2)^2 \ln (2)
                \nonumber\\&
                -280 (n.p_1)^3 n.p_2 \ln(\hat{r})
                +900 (n.p_1)^2 (n.p_2)^2 \ln(\hat{r})
                -280 n.p_1 (n.p_2)^3 \ln(\hat{r})
                \nonumber\\&
                +\frac{776}{5} (n.p_1)^2 p_1.p_2 \ln(\hat{r})
                -528 n.p_1 n.p_2 p_1.p_2 \ln(\hat{r})
                +\frac{776}{5} (n.p_2)^2 p_1.p_2 \ln(\hat{r})
                \nonumber\\&
                +56 (p_1.p_2)^2 \ln(\hat{r})
                +\frac{P_{235} m_1^3}{m_2^3}
                +
                \frac{P_{257} m_1^2}{m_2^2}
                +\frac{P_{255} m_1}{m_2}
                +\frac{P_{213} m_2}{m_1}
                +\frac{P_{110} m_2^2}{m_1^2}
                +\frac{P_{92} m_2^3}{m_1^3}
                \nonumber\\&
                -\frac{284573 n.p_1 n.p_2 p_1^2}{1800}
                +\frac{220249}{600} (n.p_2)^2 p_1^2
                +\frac{97787 p_1.p_2 p_1^2}{1800}
                +\frac{638}{15} n.p_1 n.p_2 \ln (2) p_1^2
                \nonumber\\&
                -92 (n.p_2)^2 \ln (2) p_1^2
                -\frac{574}{45} p_1.p_2 \ln (2) p_1^2
                +\frac{638}{5} n.p_1 n.p_2 \ln(\hat{r}) p_1^2
                -276 (n.p_2)^2 \ln(\hat{r}) p_1^2
                \nonumber\\&
                -\frac{574}{15} p_1.p_2 \ln(\hat{r}) p_1^2
                +\frac{220249}{600} (n.p_1)^2 p_2^2
                -\frac{284573 n.p_1 n.p_2 p_2^2}{1800}
                +\frac{97787 p_1.p_2 p_2^2}{1800}
                \nonumber\\&
                -92 (n.p_1)^2 \ln (2) p_2^2
                +\frac{638}{15} n.p_1 n.p_2 \ln (2) p_2^2
                -\frac{574}{45} p_1.p_2 \ln (2) p_2^2
                -276 (n.p_1)^2 \ln(\hat{r}) p_2^2
                \nonumber\\&
                +\frac{638}{5} n.p_1 n.p_2 \ln(\hat{r}) p_2^2
                -\frac{574}{15} p_1.p_2 \ln(\hat{r}) p_2^2
                -\frac{3685}{24} p_1^2 p_2^2
                +\frac{124}{3} \ln (2) p_1^2 p_2^2
                \nonumber\\&
                +124 \ln(\hat{r}) p_1^2 p_2^2
        \bigg] \left(\frac{G_N}{r}\right)^3 S_{\varepsilon }^2
        +\bigg[
                \pi ^2 P_{293}
                +\varepsilon  \bigg(
                        \pi ^2 P_{296}
                        +P_{195} m_1^3
                        +\frac{P_{164} m_1^4}{m_2}
                        \nonumber\\&
                        +P_{194} m_1^2 m_2
                        +P_{196} m_1 m_2^2
                        +P_{89} m_2^3
                        +\frac{P_{70} m_2^4}{m_1}
                \bigg)
                +P_{171} m_1^3
                +\frac{P_{136} m_1^4}{m_2}
                +P_{169} m_1^2 m_2
                \nonumber\\&
                +P_{175} m_1 m_2^2
                +P_{76} m_2^3
                +\frac{P_{51} m_2^4}{m_1}
        \bigg] \left(\frac{G_N}{r}\right)^4 S_{\varepsilon }^3
        +P_{41} \left(\frac{G_N}{r}\right)^5 S_{\varepsilon }^4
\bigg\}
+\tilde{S}_{\varepsilon}^5 P_{34} \left(\frac{G_N}{r}\right)^5.
\end{align}
%----------------------------------------------------------------------------------------------------------------------------
\noindent

\vspace*{5mm}
\noindent
{\bf Acknowledgment.} We thank S.~Foffa and R.~Sturani for a discussion on Ref.~\cite{Foffa:2019rdf} and
Th.~Damour, L.~Blanchet, K.~Sch\"onwald and J.~Steinhoff for different discussions. This work has been funded in part by 
EU TMR network SAGEX agreement No. 764850 (Marie Sk\l{}odowska-Curie) and COST action CA16201: Unraveling 
new physics at the LHC through the precision frontier. G.~Sch\"afer has been supported in part by Kolleg 
Mathematik Physik Berlin (KMPB). Part of the text has been typesetted using {\tt SigmaToTeX} of the 
package {\tt Sigma} \cite{SIG1,SIG2}.

{\small
%-------------------------------------------------------------------------------------

%------------------------------------------------------------------------------

\begin{thebibliography}{99}
%-------------------------------------------------------------------------------------
%
%[1]
\bibitem{TheLIGOScientific:2016pea}
  B.P.~Abbott {\it et al.} [LIGO Scientific and Virgo Collaborations],
  %``Binary Black Hole Mergers in the first Advanced LIGO Observing Run,''
  Phys.\ Rev.\ X {\bf 6} (2016) no.4,  041015
   Erratum: [Phys.\ Rev.\ X {\bf 8} (2018) no.3,  039903]
%  doi:10.1103/PhysRevX.6.041015, 10.1103/PhysRevX.8.039903
  [arXiv:1606.04856 [gr-qc]];
  %%CITATION = doi:10.1103/PhysRevX.6.041015, 10.1103/PhysRevX.8.039903;%%
%\bibitem{LIGOScientific:2018mvr}
%  B.P.~Abbott {\it et al.} [LIGO Scientific and Virgo Collaborations],
  %``GWTC-1: A Gravitational-Wave Transient Catalog of Compact Binary Mergers Observed by LIGO and Virgo during the First and Second Observing Runs,''
  Phys.\ Rev.\ X {\bf 9} (2019) no.3,  031040
%  doi:10.1103/PhysRevX.9.031040
  [arXiv:1811.12907 [astro-ph.HE]].
  %%CITATION = doi:10.1103/PhysRevX.9.031040;%%
%-------------------------------------------------------------------------------------
%
%[2]
\bibitem{Kol:2007bc}
  B.~Kol and M.~Smolkin,
  %``Non-Relativistic Gravitation: From Newton to Einstein and Back,''
  Class.\ Quant.\ Grav.\  {\bf 25} (2008) 145011
%  doi:10.1088/0264-9381/25/14/145011
  [arXiv:0712.4116 [hep-th]];\\
  %%CITATION = doi:10.1088/0264-9381/25/14/145011;%%
%\bibitem{Levi:2018nxp}
  M.~Levi,
  %``Effective Field Theories of Post-Newtonian Gravity: A comprehensive review,''
  Rep. Progr. Phys. in print
  [arXiv:1807.01699 [hep-th]].
  %%CITATION = ARXIV:1807.01699;%%
%-------------------------------------------------------------------------------------
%
%[3]
\bibitem{EFT}
W.D.~Goldberger and I.Z.~Rothstein,
%%``An effective field theory of gravity for extended objects,''
Phys.\ Rev.\ D {\bf 73} (2006) 104029
%doi:10.1103/PhysRevD.73.104029
[hep-th/0409156].
  %%CITATION = doi:10.1103/PhysRevD.73.104029;%%
%-------------------------------------------------------------------------------------
%
%[4]
\bibitem{Gilmore:2008gq}
  J.B.~Gilmore and A.~Ross,
  %``Effective field theory calculation of second post--Newtonian binary dynamics,''
  Phys.\ Rev.\ D {\bf 78} (2008) 124021
%  doi:10.1103/PhysRevD.78.124021
  [arXiv:0810.1328 [gr-qc]].
  %%CITATION = doi:10.1103/PhysRevD.78.124021;%%
%-------------------------------------------------------------------------------------
%
%[5]
\bibitem{Foffa:2011ub}
  S.~Foffa and R.~Sturani,
  %``Effective field theory calculation of conservative binary dynamics at third post--Newtonian order,''
  Phys.\ Rev.\ D {\bf 84} (2011) 044031
%  doi:10.1103/PhysRevD.84.044031
  [arXiv:1104.1122 [gr-qc]].
  %%CITATION = doi:10.1103/PhysRevD.84.044031;%%
%-------------------------------------------------------------------------------------
%
%[6]
\bibitem{Foffa:2019hrb}
  S.~Foffa, P.~Mastrolia, R.~Sturani, C.~Sturm and W.~J.~Torres Bobadilla,
  %``Static two-body potential at fifth post--Newtonian order,''
  Phys.\ Rev.\ Lett.\  {\bf 122} (2019) no.24,  241605
%  doi:10.1103/PhysRevLett.122.241605
  [arXiv:1902.10571 [gr-qc]].
  %%CITATION = doi:10.1103/PhysRevLett.122.241605;%%
%-------------------------------------------------------------------------------------
%
%[7]
\bibitem{Blumlein:2019zku}
  J.~Bl\"umlein, A.~Maier and P.~Marquard,
  %``Five-Loop Static Contribution to the Gravitational Interaction Potential of Two Point Masses,''
  Phys.\ Lett.\ B {\bf 800} (2020) 135100
%  doi:10.1016/j.physletb.2019.135100
  [arXiv:1902.11180 [gr-qc]].
  %%CITATION = doi:10.1016/j.physletb.2019.135100;%%
%-------------------------------------------------------------------------------------
%
%[8]
\bibitem{Foffa:2019rdf}
  S.~Foffa and R.~Sturani,
  %``Conservative dynamics of binary systems to fourth Post-Newtonian order in the EFT approach I: Regularized Lagrangian,''
  Phys.\ Rev.\ D {\bf 100} (2019) no.2,  024047
%  doi:10.1103/PhysRevD.100.024047
  [arXiv:1903.05113 [gr-qc]].
  %%CITATION = doi:10.1103/PhysRevD.100.024047;%%
%-------------------------------------------------------------------------------------
%
%[9]
\bibitem{Foffa:2019yfl}
  S.~Foffa, R.A.~Porto, I.~Rothstein and R.~Sturani,
  %``Conservative dynamics of binary systems to fourth Post-Newtonian order in the EFT approach II: Renormalized Lagrangian,''
  Phys.\ Rev.\ D {\bf 100} (2019) no.2,  024048
%  doi:10.1103/PhysRevD.100.024048
  [arXiv:1903.05118 [gr-qc]].
  %%CITATION = doi:10.1103/PhysRevD.100.024048;%%
%-------------------------------------------------------------------------------------
%
%[10]
\bibitem{Foffa:2019eeb}
  S.~Foffa and R.~Sturani,
  {\it Hereditary Terms at Next-To-Leading Order in Two-Body Gravitational Dynamics},
  arXiv:1907.02869 [gr-qc];\\
  %%CITATION = ARXIV:1907.02869;%%
%\bibitem{Blanchet:2019rjs}
  L.~Blanchet, S.~Foffa, F.~Larrouturou and R.~Sturani,
  {\it Logarithmic tail contributions to the energy function of circular compact binaries},
  arXiv:1912.12359 [gr-qc].
  %%CITATION = ARXIV:1912.12359;%%
%-------------------------------------------------------------------------------------
%
%[11]
\bibitem{Damour:2014jta}
  T.~Damour, P.~Jaranowski and G.~Sch\"afer,
  %``Nonlocal-in-time action for the fourth post--Newtonian conservative dynamics of two-body systems,''
  Phys.\ Rev.\ D {\bf 89} (2014) no.6,  064058
%  doi:10.1103/PhysRevD.89.064058
  [arXiv:1401.4548 [gr-qc]].
  %%CITATION = doi:10.1103/PhysRevD.89.064058;%%
%-------------------------------------------------------------------------------------
%
%[12]
\bibitem{Bernard:2017ktp}
  L.~Bernard, L.~Blanchet, G.~Faye and T.~Marchand,
  %``Center-of-Mass Equations of Motion and Conserved Integrals of Compact Binary Systems at the Fourth Post-Newtonian Order,''
  Phys.\ Rev.\ D {\bf 97} (2018) no.4,  044037
%  doi:10.1103/PhysRevD.97.044037
  [arXiv:1711.00283 [gr-qc]].
  %%CITATION = doi:10.1103/PhysRevD.97.044037;%%
%-------------------------------------------------------------------------------------
%
%[13]
\bibitem{Bini:2019nra}
  D.~Bini, T.~Damour and A.~Geralico,
  %``Novel approach to binary dynamics: application to the fifth post--Newtonian level,''
  Phys.\ Rev.\ Lett.\  {\bf 123} (2019) no.23,  231104
%  doi:10.1103/PhysRevLett.123.231104
  [arXiv:1909.02375 [gr-qc]] and references therein.
  %%CITATION = doi:10.1103/PhysRevLett.123.231104;%%
%-------------------------------------------------------------------------------------
%
%[14]
\bibitem{CLAUSIUS}
R.~Clausius, Ann. Phys. (Leipzig) {\bf 217} (1870) 124--130;\\
H.~Stephani and G.~Kluge, {\sf Grundlagen der theoretischen Mechanik}
(DVW, Berlin, 1975).
%-------------------------------------------------------------------------------------
%
%[15]
\bibitem{Bern:2019crd}
  Z.~Bern, C.~Cheung, R.~Roiban, C.H.~Shen, M.P.~Solon and M.~Zeng,
  %``Black Hole Binary Dynamics from the Double Copy and Effective Theory,''
  JHEP {\bf 1910} (2019) 206
%  doi:10.1007/JHEP10(2019)206
  [arXiv:1908.01493 [hep-th]];
  %%CITATION = doi:10.1007/JHEP10(2019)206;%%
%\bibitem{Bern:2019nnu}
%  Z.~Bern, C.~Cheung, R.~Roiban, C.H.~Shen, M.P.~Solon and M.~Zeng,
  %``Scattering Amplitudes and the Conservative Hamiltonian for Binary Systems at Third Post-Minkowskian Order,''
  Phys.\ Rev.\ Lett.\  {\bf 122} (2019) no.20,  201603
%  doi:10.1103/PhysRevLett.122.201603
  [arXiv:1901.04424 [hep-th]].
  %%CITATION = doi:10.1103/PhysRevLett.122.201603;%%
%-------------------------------------------------------------------------------------
%
%[16]
\bibitem{Damour:2019lcq}
  T.~Damour,
  {\it Classical and Quantum Scattering in Post-Minkowskian Gravity},
  arXiv:1912.02139 [gr-qc].
  %%CITATION = ARXIV:1912.02139;%%
%-------------------------------------------------------------------------------------
%
%[17]
\bibitem{PM}
%\bibitem{Cheung:2018wkq}
  C.~Cheung, I.Z.~Rothstein and M.P.~Solon,
  %``From Scattering Amplitudes to Classical Potentials in the Post-Minkowskian Expansion,''
  Phys.\ Rev.\ Lett.\  {\bf 121} (2018) no.25,  251101
%  doi:10.1103/PhysRevLett.121.251101
  [arXiv:1808.02489 [hep-th]];\\
  %%CITATION = doi:10.1103/PhysRevLett.121.251101;%%
%\bibitem{DAM1}
%\bibitem{Damour:2017zjx}
%  T.~Damour,
  %``High-energy gravitational scattering and the general relativistic two-body problem,''
  Phys.\ Rev.\ D {\bf 97} (2018) no.4,  044038
%  doi:10.1103/PhysRevD.97.044038
  [arXiv:1710.10599 [gr-qc]];\\
  %%CITATION = doi:10.1103/PhysRevD.97.044038;%%
%------------------------------------------------------------------------------
%\bibitem{Damour:2016gwp}
  T.~Damour,
  %``Gravitational scattering, post-Minkowskian approximation and Effective One-Body theory,''
  Phys.\ Rev.\ D {\bf 94} (2016) no.10,  104015
%  doi:10.1103/PhysRevD.94.104015
  [arXiv:1609.00354 [gr-qc]];
  %%CITATION = doi:10.1103/PhysRevD.94.104015;%%
%\bibitem{Damour:2017zjx}
%  T.~Damour,
  %``High-energy gravitational scattering and the general relativistic two-body problem,''
  Phys.\ Rev.\ D {\bf 97} (2018) no.4,  044038
%  doi:10.1103/PhysRevD.97.044038
  [arXiv:1710.10599 [gr-qc]];\\
  %%CITATION = doi:10.1103/PhysRevD.97.044038;%%
%\bibitem{Bjerrum-Bohr:2018xdl}
  N.E.J.~Bjerrum-Bohr, P.H.~Damgaard, G.~Festuccia, L.~Plant\'e and P.~Vanhove,
  %``General Relativity from Scattering Amplitudes,''
  Phys.\ Rev.\ Lett.\  {\bf 121} (2018) no.17,  171601
%  doi:10.1103/PhysRevLett.121.171601
  [arXiv:1806.04920 [hep-th]];\\
  %%CITATION = doi:10.1103/PhysRevLett.121.171601;%%
%\bibitem{KoemansCollado:2019ggb}
  A.~Koemans Collado, P.~Di Vecchia and R.~Russo,
  %``Revisiting the second post-Minkowskian eikonal and the dynamics of binary black holes,''
  Phys.\ Rev.\ D {\bf 100} (2019) no.6,  066028
%  doi:10.1103/PhysRevD.100.066028
  [arXiv:1904.02667 [hep-th]];\\
  %%CITATION = doi:10.1103/PhysRevD.100.066028;%%
%\bibitem{Cristofoli:2019neg}
  A.~Cristofoli, N.E.J.~Bjerrum-Bohr, P.H.~Damgaard and P.~Vanhove,
  %``On Post-Minkowskian Hamiltonians in General Relativity,''
  Phys.\ Rev.\ D {\bf 100} (2019) no.8,  084040
%  doi:10.1103/PhysRevD.100.084040
  [arXiv:1906.01579 [hep-th]];\\
  %%CITATION = doi:10.1103/PhysRevD.100.084040;%%
%\bibitem{Bjerrum-Bohr:2019kec}
  N.E.J.~Bjerrum-Bohr, A.~Cristofoli and P.H.~Damgaard,
  %``Post-Minkowskian Scattering Angle in Einstein Gravity,''
  arXiv:1910.09366 [hep-th];\\
%\bibitem{Blumlein:2019bqq}
  J.~Bl\"umlein, A.~Maier, P.~Marquard, G.~Sch\"afer and C.~Schneider,
  %``From Momentum Expansions to Post-Minkowskian Hamiltonians by Computer Algebra Algorithms,''
  Phys.\ Lett.\ B {\bf 801} (2020) 135157
%  doi:10.1016/j.physletb.2019.135157
  [arXiv:1911.04411 [gr-qc]].
  %%CITATION = doi:10.1016/j.physletb.2019.135157;%%
%-------------------------------------------------------------------------------------
%
%[18]
\bibitem{Schafer:2018kuf}
  G.~Sch\"afer and P.~Jaranowski,
  %``Hamiltonian formulation of general relativity and post--Newtonian dynamics of compact binaries,''
  Living Rev.\ Rel.\  {\bf 21} (2018) no.1,  7, 1--117
%  doi:10.1007/s41114-018-0016-5
  [arXiv:1805.07240 [gr-qc]].
  %%CITATION = doi:10.1007/s41114-018-0016-5;%%
%-------------------------------------------------------------------------------------
%
%[19]
\bibitem{Nogueira:1991ex}
  P.~Nogueira,
  %``Automatic Feynman graph generation,''
  J.\ Comput.\ Phys.\  {\bf 105} (1993) 279--289.
%  doi:10.1006/jcph.1993.1074
  %%CITATION = doi:10.1006/jcph.1993.1074;%%
%-------------------------------------------------------------------------------------
%
%[20]
\bibitem{FORM}
%\bibitem{Vermaseren:2000nd}
  J.A.M.~Vermaseren, 
  {\it New features of FORM},
  math-ph/0010025;\\
  %%CITATION = MATH-PH/0010025;%%
%\bibitem{Tentyukov:2007mu}
  M.~Tentyukov and J.A.M.~Vermaseren,
  %``The Multithreaded version of FORM,''
  Comput.\ Phys.\ Commun.\  {\bf 181} (2010) 1419--1427
%  doi:10.1016/j.cpc.2010.04.009
  [hep-ph/0702279].
  %%CITATION = doi:10.1016/j.cpc.2010.04.009;%%
%------------------------------------------------------------------------------
%
%[21]
\bibitem{CRUSHER}
P.~Marquard and D.~Seidel, {\it The {\tt Crusher} algorithm}, unpublished.
%------------------------------------------------------------------------------
%
%[22]
\bibitem{Foffa:2012rn}
  S.~Foffa and R.~Sturani,
  %``Dynamics of the gravitational two-body problem at fourth post--Newtonian order and at quadratic order in the Newton constant,''
  Phys.\ Rev.\ D {\bf 87} (2013) no.6,  064011
%  doi:10.1103/PhysRevD.87.064011
  [arXiv:1206.7087 [gr-qc]].
  %%CITATION = doi:10.1103/PhysRevD.87.064011;%%
%------------------------------------------------------------------------------
%
%[23]
\bibitem{SCHMUTZER}
E.~Schmutzer, {\sf Grundprinzipien der klassischen Mechanik und der klassischen Feldtheorie}, (DVW, Berlin, 1973).
%-------------------------------------------------------------------------------------
%
%[24]
\bibitem{Porto:2017dgs}
  R.A.~Porto and I.Z.~Rothstein,
  %``Apparent ambiguities in the post--Newtonian expansion for binary systems,''
  Phys.\ Rev.\ D {\bf 96} (2017) no.2,  024062
%  doi:10.1103/PhysRevD.96.024062
  [arXiv:1703.06433 [gr-qc]].
  %%CITATION = doi:10.1103/PhysRevD.96.024062;%%
%------------------------------------------------------------------------------
%
%[25]
\bibitem{Damour:1985mt} 
  T.~Damour and G.~Sch\"afer,
  %``Lagrangians forn point masses at the second post--Newtonian approximation of general relativity,''
  Gen.\ Rel.\ Grav.\  {\bf 17} (1985) 879--905. 
  %%CITATION = doi:10.1007/BF00773685;%%
%------------------------------------------------------------------------------------- 
%
%[26]
\bibitem{DW}
B.S. DeWitt, {\it Dynamical Theory of Groups and Fields}  in {\sf Relativiy, Groups and Topology},
Eds.~C.~DeWitt and B.~DeWitt, (Gordon and Breach, New York, 1964), Eq.~(18.1).
%-------------------------------------------------------------------------------------
%
%[27]
\bibitem{Damour:1990jh}
  T.~Damour and G.~Sch\"afer,
  %``Redefinition of position variables and the reduction of higher order Lagrangians,''
  J.\ Math.\ Phys.\  {\bf 32} (1991) 127--134.
%  doi:10.1063/1.529135
  %%CITATION = doi:10.1063/1.529135;%%
%-------------------------------------------------------------------------------------
%
%[28]
\bibitem{Marchand:2017pir}
  T.~Marchand, L.~Bernard, L.~Blanchet and G.~Faye,
  %``Ambiguity-Free Completion of the Equations of Motion of Compact Binary Systems at the Fourth Post-Newtonian Order,''
  Phys.\ Rev.\ D {\bf 97} (2018) no.4,  044023
%  doi:10.1103/PhysRevD.97.044023
  [arXiv:1707.09289 [gr-qc]].
  %%CITATION = doi:10.1103/PhysRevD.97.044023;%%
%-------------------------------------------------------------------------------------
%
%[29]
\bibitem{GS:1990}
G.~Sch\"afer, Astron. Nachr. {\bf 311} (1990) 213--217.
%-------------------------------------------------------------------------------------
% 
%[30]
\bibitem{Damour:2016abl}
  T.~Damour, P.~Jaranowski and G.~Sch\"afer,
  %``Conservative dynamics of two-body systems at the fourth post--Newtonian approximation of general relativity,''
  Phys.\ Rev.\ D {\bf 93} (2016) no.8,  084014
%  doi:10.1103/PhysRevD.93.084014
  [arXiv:1601.01283 [gr-qc]].
  %%CITATION = doi:10.1103/PhysRevD.93.084014;%%
%------------------------------------------------------------------------------------- 
%
%[31]
\bibitem{Blanchet:1987wq}
  L.~Blanchet and T.~Damour,
  %``Tail Transported Temporal Correlations in the Dynamics of a Gravitating System,''
  Phys.\ Rev.\ D {\bf 37} (1988) 1410--1435.
%  doi:10.1103/PhysRevD.37.1410
  %%CITATION = doi:10.1103/PhysRevD.37.1410;%%
%------------------------------------------------------------------------------------- 
%
%[32]
\bibitem{Foffa:2011np}
  S.~Foffa and R.~Sturani,
  %``Tail terms in gravitational radiation reaction via effective field theory,''
  Phys.\ Rev.\ D {\bf 87} (2013) no.4,  044056
%  doi:10.1103/PhysRevD.87.044056
  [arXiv:1111.5488 [gr-qc]].
  %%CITATION = doi:10.1103/PhysRevD.87.044056;%%
%------------------------------------------------------------------------------------- 
%
%[33]
\bibitem{Galley:2015kus}
  C.~R.~Galley, A.~K.~Leibovich, R.~A.~Porto and A.~Ross,
  %``Tail effect in gravitational radiation reaction: Time nonlocality and renormalization group evolution,''
  Phys.\ Rev.\ D {\bf 93} (2016) 124010
%  doi:10.1103/PhysRevD.93.124010
  [arXiv:1511.07379 [gr-qc]].
  %%CITATION = doi:10.1103/PhysRevD.93.124010;%%
%------------------------------------------------------------------------------------- 
% 
%[34]
\bibitem{Bernard:2017bvn}
  L.~Bernard, L.~Blanchet, A.~Boh\'{e}, G.~Faye and S.~Marsat,
  %``Dimensional regularization of the IR divergences in the Fokker action of point-particle binaries at the fourth post--Newtonian order,''
  Phys.\ Rev.\ D {\bf 96} (2017) no.10,  104043
%  doi:10.1103/PhysRevD.96.104043
  [arXiv:1706.08480 [gr-qc]].
  %%CITATION = doi:10.1103/PhysRevD.96.104043;%%
%------------------------------------------------------------------------------------- 
%
%[35]
\bibitem{Jaranowski:2015lha}
  P.~Jaranowski and G.~Sch\"afer,
  %``Derivation of local-in-time fourth post--Newtonian ADM Hamiltonian for spinless compact binaries,''
  Phys.\ Rev.\ D {\bf 92} (2015) no.12,  124043
%  doi:10.1103/PhysRevD.92.124043
  [arXiv:1508.01016 [gr-qc]].
  %%CITATION = doi:10.1103/PhysRevD.92.124043;%%
%-------------------------------------------------------------------------------------
%
%[36]
\bibitem{LUETZEN}
J.~L\"utzen, {\sf The Prehistory of the Theory of Distributions}, (Springer, New York, 1982).
%-------------------------------------------------------------------------------------
%
%[37]
\bibitem{HADAMARD}
J.~Hadamard, {\sf Le probl\`{e}me de Cauchy et les \'{e}quations aux d\'{e}riv\'{e}es partielles 
lin\'{e}aires hyperboliques}, (Hermann et Cie., Paris,  1932);\\
J.~Naas, H.L.~Schmid, {\sf Mathematisches W\"orterbuch}, (Akademie Verlag, Berlin, 1961), Vol.~I, p.~448.
%-------------------------------------------------------------------------------------
%
%[38]
\bibitem{VLADIMIROV}
V.S.~Vladimirov, {\sf Gleichungen der mathematischen Physik}, (DVW, Berlin, 1972).
%-------------------------------------------------------------------------------------
% 
%[39]
\bibitem{WEINB}
S.~Weinberg, {Gravitation and Cosmology, Principles and Applications of the General Theory of Relativity}, (J. Wiley \& 
Sons, Hoboken, NJ, 1972).
%------------------------------------------------------------------------------
%
%[40]
\bibitem{SCHMUTZER1}
E.~Schmutzer, {\sf Relativistische Physik}, (Teubner, Leipzig, 1968).
%-------------------------------------------------------------------------------------
%
%[41]
\bibitem{Symanzik:1970rt}
  K.~Symanzik,
  %``Small distance behavior in field theory and power counting,''
  Commun.\ Math.\ Phys.\  {\bf 18} (1970) 227--246.
%  doi:10.1007/BF01649434
  %%CITATION = doi:10.1007/BF01649434;%%
%------------------------------------------------------------------------------
%
%[42]
\bibitem{Callan:1970yg}
  C.G.~Callan, Jr.,
  %``Broken scale invariance in scalar field theory,''
  Phys.\ Rev.\ D {\bf 2} (1970) 1541--1547.
%  doi:10.1103/PhysRevD.2.1541
  %%CITATION = doi:10.1103/PhysRevD.2.1541;%%
%------------------------------------------------------------------------------
%
%[43]
\bibitem{Alvey:2019ctk}
  J.~Alvey, N.~Sabti, M.~Escudero and M.~Fairbairn,
  {\it Improved BBN Constraints on the Variation of the Gravitational Constant},
  arXiv:1910.10730 [astro-ph.CO].
  %%CITATION = ARXIV:1910.10730;%%   }.
%------------------------------------------------------------------------------
%
%[44]
\bibitem{MITTELSTAEDT}
P.~Mittelst\"adt, {\sf Klassische Mechanik}, 2nd Ed., BI Vol.~500, (BI Wissenschaftsverlag, Mannheim, 1995).
%------------------------------------------------------------------------------
k%
%[45]
\bibitem{GROEBNER}
W.~Gr\"obner, {\sf Die Lie-Reihen und ihre Anwendungen}, (DVW, Berlin, 1960).
%------------------------------------------------------------------------------
%
%[46]
\bibitem{VINTI}
J.P.~Vinti,  {\sf Orbital and Celestial Mechanics. Progress in Astronautics and Aeronautics}, {\bf 177}, 
(1998) American Institute of Aeronautics and Astronautics, Reston, VA, 
G.J~Der and N.L~Bonavito (Eds.).
%------------------------------------------------------------------------------
%
%[47]
\bibitem{JACOBI1}
C.G.~Jacobi, {\sf Vorlesungen \"uber analytische Mechanik}, nach einer Mitschrift von Wilhelm Scheibner, 
Friedrich Wilhelms Universit\"at Berlin 1847/48, Ed.~H.~ Pulte, (Vieweg, Wiesbaden, 1996).
%------------------------------------------------------------------------------
%
%[48]
\bibitem{KILLING}
W.~Killing, Math. Ann. {\bf 31} (1888) 252--290; {\bf 33} (1889) 1--48; {\bf 34} (1889) 57--122;
{\bf 36} (1890) 161--189.
%------------------------------------------------------------------------------
%
%[49]
\bibitem{CARTAN}
E.~Cartan, {\it Sur la structure des groupes de transformations finis et continus}, Premi\`ere Th\`ese, L' \'Ecole Normale 
Sup\'erior, (Libraire Nony \& C, Paris, 1894).
%------------------------------------------------------------------------------
%
%[50]
\bibitem{HAMJAC}
W.R.~Hamilton, Phil. Trans. Roy. Soc. {\bf 124} part II (1834) 247--308;
Phil. Trans. Roy. Soc. of London {\bf 125} (1835) 95--144;\\ 
C.G.~Jacobi, Gesammelte Werke, Supplement, Ed.~E.~Lottner, {\sf Vorlesungen \"uber Mechanik}, (G.~Reimer, Berlin, 1884) gehalten 
1842/43 [Mitschrift von C.W.~Borchart].
%------------------------------------------------------------------------------
%
%[51]
\bibitem{POINEIN}
H. Poincar\'e,  {\it Der gegenw\"artige Zustand und die Zukunft der mathematischen Physik}, In: {\sf Der Wert der 
Wissenschaft}, Chap.~7--8, (Teubner, Leipzig, 1906), authorized translation by H.~Weber; [{\sf La valeur de la science}, 
(Flammarion, Paris, 1904, Biblioth\`eque de la philosophie scientifique)]; {\sf Wissenschaft und Hypothese}, Chap.~5,
German translation: F. and  L. Lindemann, (Teubner, Leipzig, 1904) [{\sf La Science et l'Hypoth\'ese}, (Flammarion, Paris, 
1902, Biblioth\`eque de la philosophie scientifique)];
\\
A. Einstein, Ann. Phys. {\bf 322} (1905) 89--921.
%-------------------------------------------------------------------------------------
%
%[52]
\bibitem{Buonanno:1998gg}
  A.~Buonanno and T.~Damour,
  %``Effective one-body approach to general relativistic two-body dynamics,''
  Phys.\ Rev.\ D {\bf 59} (1999) 084006
%  doi:10.1103/PhysRevD.59.084006
  [gr-qc/9811091];
  %%CITATION = doi:10.1103/PhysRevD.59.084006;%%
%-------------------------------------------------------------------------------------
%
%[52]
\bibitem{Buonanno:2000ef}   
  A.~Buonanno and T.~Damour,
  %``Transition from inspiral to plunge in binary black hole coalescences,''
  Phys.\ Rev.\ D {\bf 62} (2000) 064015
%  doi:10.1103/PhysRevD.62.064015
  [gr-qc/0001013].
  %%CITATION = doi:10.1103/PhysRevD.62.064015;%%
%------------------------------------------------------------------------------
%
%[53]
\bibitem{THIRD}
%\bibitem{Damour:2000we}
  T.~Damour, P.~Jaranowski and G.~Sch\"afer,
  %``On the determination of the last stable orbit for circular general relativistic binaries at the third postNewtonian approximation,''
  Phys.\ Rev.\ D {\bf 62} (2000) 084011
%  doi:10.1103/PhysRevD.62.084011
  [gr-qc/0005034].
  %%CITATION = doi:10.1103/PhysRevD.62.084011;%%
%------------------------------------------------------------------------------
%
%[54]
\bibitem{Damour:2015isa}
  T.~Damour, P.~Jaranowski and G.~Sch\"afer,
  %``Fourth post--Newtonian effective one-body dynamics,''
  Phys.\ Rev.\ D {\bf 91} (2015) no.8,  084024
%  doi:10.1103/PhysRevD.91.084024
  [arXiv:1502.07245 [gr-qc]].
  %%CITATION = doi:10.1103/PhysRevD.91.084024;%%
%------------------------------------------------------------------------------
%
%[55]
\bibitem{EXC}
%\bibitem{Damour:2009sm}
  T.~Damour,
  %``Gravitational Self Force in a Schwarzschild Background and the Effective One Body Formalism,''
  Phys.\ Rev.\ D {\bf 81} (2010) 024017
%  doi:10.1103/PhysRevD.81.024017
  [arXiv:0910.5533 [gr-qc]];\\
  %%CITATION = doi:10.1103/PhysRevD.81.024017;%%
%\bibitem{LeTiec:2011bk}
  A.~Le Tiec, A.~H.~Mroue, L.~Barack, A.~Buonanno, H.~P.~Pfeiffer, N.~Sago and A.~Taracchini,
  %``Periastron Advance in Black Hole Binaries,''
  Phys.\ Rev.\ Lett.\  {\bf 107} (2011) 141101
%  doi:10.1103/PhysRevLett.107.141101
  [arXiv:1106.3278 [gr-qc]];\\
  %%CITATION = doi:10.1103/PhysRevLett.107.141101;%%
%\bibitem{Bini:2015bfb}
  D.~Bini, T.~Damour and A.~Geralico,
  %``Confirming and improving post-Newtonian and effective-one-body results from self-force computations along eccentric orbits around a Schwarzschild black 
hole,''
  Phys.\ Rev.\ D {\bf 93} (2016) no.6,  064023
%  doi:10.1103/PhysRevD.93.064023
  [arXiv:1511.04533 [gr-qc]];\\
  %%CITATION = doi:10.1103/PhysRevD.93.064023;%%
%\bibitem{Hopper:2015icj}
  S.~Hopper, C.~Kavanagh and A.C.~Ottewill,
  %``Analytic self-force calculations in the post-Newtonian regime: eccentric orbits on a Schwarzschild background,''
  Phys.\ Rev.\ D {\bf 93} (2016) no.4,  044010
%  doi:10.1103/PhysRevD.93.044010
  [arXiv:1512.01556 [gr-qc]];\\
  %%CITATION = doi:10.1103/PhysRevD.93.044010;%%
%\bibitem{Bini:2016qtx}
  D.~Bini, T.~Damour and A.~Geralico,
  %``New gravitational self-force analytical results for eccentric orbits around a Schwarzschild black hole,''
  Phys.\ Rev.\ D {\bf 93} (2016) no.10,  104017
%  doi:10.1103/PhysRevD.93.104017
  [arXiv:1601.02988 [gr-qc]].
  %%CITATION = doi:10.1103/PhysRevD.93.104017;%%
%-------------------------------------------------------------------------------------
%
%[56]
\bibitem{Antonelli:2019ytb}
  A.~Antonelli, A.~Buonanno, J.~Steinhoff, M.~van de Meent and J.~Vines,
  %``Energetics of two-body Hamiltonians in post-Minkowskian gravity,''
  Phys.\ Rev.\ D {\bf 99} (2019) no.10,  104004
%  doi:10.1103/PhysRevD.99.104004
  [arXiv:1901.07102 [gr-qc]];\\
  %%CITATION = doi:10.1103/PhysRevD.99.104004;%%
%\bibitem{Buonanno:2002ft}
  A.~Buonanno, Y.B.~Chen and M.~Vallisneri,
  %``Detection template families for gravitational waves from the final stages of binary--black-hole inspirals: Nonspinning case,''
  Phys.\ Rev.\ D {\bf 67} (2003) 024016
  Erratum: [Phys.\ Rev.\ D {\bf 74} (2006) 029903]
%  doi:10.1103/PhysRevD.67.024016, 10.1103/PhysRevD.74.029903
  [gr-qc/0205122].
  %%CITATION = doi:10.1103/PhysRevD.67.024016, 10.1103/PhysRevD.74.029903;%%
%-------------------------------------------------------------------------------------
%
%[57]
\bibitem{Jaranowski:2013lca}
  P.~Jaranowski and G.~Sch\"afer,
  %``Dimensional regularization of local singularities in the 4th post--Newtonian two-point-mass Hamiltonian,''
  Phys.\ Rev.\ D {\bf 87} (2013) 081503
%  doi:10.1103/PhysRevD.87.081503
  [arXiv:1303.3225 [gr-qc]].
  %%CITATION = doi:10.1103/PhysRevD.87.081503;%%
%-------------------------------------------------------------------------------------
%
%[58]
\bibitem{MERG}
%\bibitem{Damour:2012ky}
  T.~Damour, A.~Nagar and S.~Bernuzzi,
  %``Improved effective-one-body description of coalescing nonspinning black-hole binaries and its numerical-relativity completion,''
  Phys.\ Rev.\ D {\bf 87} (2013) no.8,  084035
%  doi:10.1103/PhysRevD.87.084035
  [arXiv:1212.4357 [gr-qc]];\\
  %%CITATION = doi:10.1103/PhysRevD.87.084035;%%
%\bibitem{Damour:2012mv}
  T.~Damour,
  %``The General Relativistic Two Body Problem and the Effective One Body Formalism,''
  Fundam.\ Theor.\ Phys.\  {\bf 177} (2014) 111--145
%  doi:10.1007/978-3-319-06349-2_5
  [arXiv:1212.3169 [gr-qc]];\\
  %%CITATION = doi:10.1007/978-3-319-06349-2_5;%%
%\bibitem{Taracchini:2013rva}
  A.~Taracchini {\it et al.},
  %``Effective-one-body model for black-hole binaries with generic mass ratios and spins,''
  Phys.\ Rev.\ D {\bf 89} (2014) no.6,  061502
  doi:10.1103/PhysRevD.89.061502
  [arXiv:1311.2544 [gr-qc]].
  %%CITATION = doi:10.1103/PhysRevD.89.061502;%%
%------------------------------------------------------------------------------
%
%[59]
\bibitem{JK}
P.~Jaranowski and A.~Krolak, {\sf Analysis of Gravitational-Wave Data}, Cambridge Monographs on Particle Physics, 
Nuclear Physics and Cosmology, Vol.~29, (Cambridge University Press, Cambridge, 2009).
%------------------------------------------------------------------------------
%
%[60]
\bibitem{Blanchet:2019zlt}
  L.~Blanchet,
  %``Analyzing Gravitational Waves with General Relativity,''
  Comptes Rendus Physique {\bf 20} (2019) 507--520
%  doi:10.1016/j.crhy.2019.02.004
  [arXiv:1902.09801 [gr-qc]];
  %%CITATION = doi:10.1016/j.crhy.2019.02.004;%%
L.~Blanchet, in: {\tt https://indico.desy.de/indico/event/23129/other-view?view=standard}
%------------------------------------------------------------------------------
%
%[61]
\bibitem{Bernard:2015njp}
  L.~Bernard, L.~Blanchet, A.~Boh\'{e}, G.~Faye and S.~Marsat,
  %``Fokker action of nonspinning compact binaries at the fourth post--Newtonian approximation,''
  Phys.\ Rev.\ D {\bf 93} (2016) no.8,  084037
%  doi:10.1103/PhysRevD.93.084037
  [arXiv:1512.02876 [gr-qc]].
  %%CITATION = doi:10.1103/PhysRevD.93.084037;%%
%------------------------------------------------------------------------------
%
%[62]
\bibitem{FOKKER}
K.~Schwarzschild, Nachr. Akad. Wiss. G\"ottingen II Math.-Physik. Kl. {\bf 2a} (1903), 126--131, 132--141, 
245--278;\\
H.~Tetrode, Z. Physik {\bf 10} (1922) 317--328;\\
A.D.~Fokker, Z. Physik {\bf 58} (1929) 386--393; Physica {\bf 9} (1929) 33--42; {\bf 12} (1932) 145--152;\\
J.A.~Wheeler and R.P.~Feynman, Rev. Mod. Phys. {\bf 17} (1945) 157--181; Rev. Mod. Phys. {\bf 21} (1949) 425--433;\\
R.P.~Feynman, Phys. Rev. {\bf 74} (1948) 939--946;\\
A.~Schild, Ann. Phys (NY) {\bf 93} (1975) 881--115.
%------------------------------------------------------------------------------
%
%[63]
\bibitem{SIG1}
C.~Schneider, { S\'em.~Lothar. Combin.\/} {\bf 56} (2007) 1--36.
%-------------------------------------------------------------------------------------
%
%[64]
\bibitem{SIG2}
 C.~Schneider, in: {\sf Computer Algebra in Quantum Field Theory: Integration,
  Summation and Special Functions}, Texts and Monographs in Symbolic
  Computation eds. C.~Schneider and J.~Bl{\"u}mlein (Springer, Wien, 2013),
  325--360 [arXiv:1304.4134 [cs.SC]].
%-------------------------------------------------------------------------------------
\end{thebibliography}
\end{document}